\documentclass[ twocolumn, 10pt]{aastex631}
\usepackage{graphicx}
\usepackage[normalem]{ulem}
\usepackage{ulem}

\usepackage{multirow}
\usepackage{url}
\usepackage{natbib}

\defcitealias{cosep}{COSEP}
\defcitealias{lsst}{I19}

\newcommand{\fom}{\ensuremath{FoM}}
\newcommand{\maf}{\texttt{MAF}}

\newcommand{\eg}{\emph{e.g.}}
\newcommand{\ie}{\emph{i.e.}}
\newcommand{\opsim}{\texttt{OpSim}}
\newcommand{\ovfive}{\texttt{OpSim~v1.5}}
\newcommand{\ovseven}{\texttt{OpSim~v1.7}}
\newcommand{\ovsevenone}{\texttt{OpSim~v1.7.1}}

\begin{document}


\title{Preparing to discover the unknown with Rubin LSST - I: Time domain}
\correspondingauthor{Xiaolong Li}
\email{lixl@udel.edu}

\author[0000-0002-0514-5650]{Xiaolong Li}
\affiliation{Department of Physics and Astronomy, University of Delaware, Newark, DE 19716, USA}

\author[0000-0003-2132-3610]{Fabio Ragosta }
\affiliation{INAF and University of Naples "Federico II", via Cinthia 9, 80126 Napoli, Italy}

\author[0000-0002-2577-8885]{William I. Clarkson}
\affiliation{Department of Natural Sciences, University of Michigan - Dearborn, 4901 Evergreen Road, Dearborn, MI 48128, USA}

\author[0000-0002-8576-1487]{Federica B. Bianco}
\affiliation{Department of Physics and Astronomy, University of Delaware, Newark, DE 19716, USA}
\affiliation{Joseph R. Biden, Jr.,  School of Public Policy and Administration, University of Delaware, Newark, DE 19717 USA}
\affiliation{Data Science Institute, University of Delaware, Newark, DE 19717 USA}
\affiliation{CUSP: Center for Urban Science and Progress, New York University, Brooklyn, NY 11201 USA}
\date{\today}

\begin{abstract}
Perhaps the most exciting promise of the Rubin Observatory Legacy Survey of Space \& Time (LSST) is its capability to discover phenomena never before seen or predicted from theory: true astrophysical novelties, but the ability of LSST to make these discoveries will depend on the survey strategy. Evaluating candidate strategies for true novelties is a challenge both practically and conceptually: unlike traditional astrophysical tracers like supernovae or exoplanets, for anomalous objects the template signal is by definition unknown. We present our approach to solve this problem, by assessing survey completeness in a phase space defined by object color, flux (and their evolution), and considering the volume explored by integrating metrics within this space with the observation depth, survey footprint, and stellar density. With these metrics, we explore recent simulations of the Rubin LSST observing strategy across the entire observed footprint and in specific regions in the Local Volume: the Galactic Plane and Magellanic Clouds. Under our metrics, observing strategies with greater diversity of exposures and time gaps tend to be more sensitive to genuinely new phenomena, particularly over time-gap ranges left relatively unexplored by previous surveys. To assist the community, we have made all the tools developed publicly available. Extension of the scheme to include proper motions and the detection of associations or populations of interest, will be communicated in paper II of this series. This paper was written with the support of the Vera C. Rubin LSST Transients and Variable Stars and Stars, Milky Way, Local Volume Science Collaborations.

\end{abstract}


\section{Introduction} \label{sec:intro}

The Rubin Observatory Legacy Survey of Space and Time (hereafter LSST)
is an ambitious project that promises to monitor the entire southern hemisphere sky over a continuous ten-year interval starting in 2024. It will deliver high sensitivity, high {(seeing-limited) spatial} resolution, and high temporal cadence (
$\geq$ 1 image per night,  $\sim$ few days repeat on each field). While other surveys have stretched into one or two directions in this feature space, delivering observations at high cadence over small fields of view (\emph{e.g.} SNLS: \citealt{snls}) or monitoring large fields of view but at low spatial resolution (\emph{e.g.} ASAS-SN: \citealt{asas-sn}), {the combination of} high spatial resolution, high cadence, and high sensitivity places the Rubin LSST  in a unique position to contribute to nearly all fields of astronomy with an unprecedentedly rich data-set.

Perhaps the most exciting promise of  Rubin LSST is thus its potential to discover as-yet unknown phenomena. This work focuses on assessing the potential of LSST to discover ``true novelties'': phenomena that have neither been observed, nor predicted, under different choices of observing strategy.


The Rubin LSST observing strategy is designed to accomplish several science goals within four science themes: (1) Probing dark energy and dark matter;
(2) Taking an inventory of the solar system;
(3) Exploring the transient optical sky;
(4) Mapping the Milky Way. These diverse goals lead to strict interlocking constraints including requirements on image quality, depth --- single visit depth and number of visits per field --- filters system, and total sky coverage. A detailed description of the science drivers and technical requirements can be found in \citet[][hereafter I19]{lsst}.\footnote{For the Science Requirements Document (SRD) itself, see \citet{lsstSRD}.} 

\begin{figure*}
\centering
\includegraphics[scale=0.7]{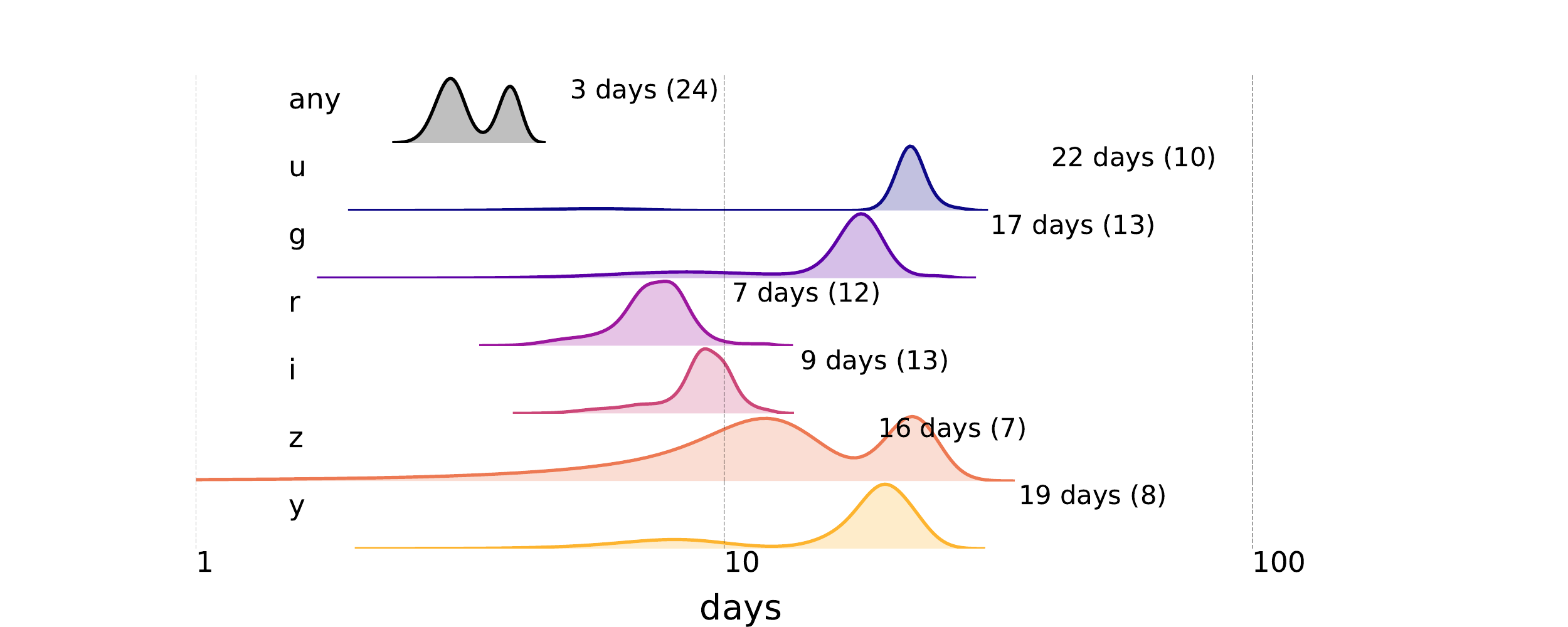}
\caption{Distribution of median time gaps between observations not in the same night for 86 simulations of the Rubin LSST Wide Fast Deep Survey \ovfive (see \autoref{ss:opsim}). 
Here, observations are spatially grouped by \texttt{healpixel} using resolution parameter \texttt{NSIDE}=64, with pixel area 0.84 square degrees 
The top row shows the distribution of gaps between observations of the same field in \emph{any} filter, the rows beneath show the time gaps between observations in the same filter, as labelled. The distributions are normalized by peak height, and the modal value over the 10 year survey simulation across all \opsim s is indicated to the right of each distribution, along with the number (in parentheses) of observations in a 1.5 day-wide bin around the peak value. The distributions are smoothed via Kernel Density Estimation (using the \texttt{Python Seaborn} package with default settings). While all these \opsim s fulfill the requirements in the SRD \citep{lsstSRD} the distributions still show a significant differences. See \autoref{sec:intro}.}
\label{fig:tgaps_example}
\end{figure*}

While the survey strategy (and indeed facility design) is thus mostly specified by the main science goals, these constraints still allow for a significant flexibility in the details. For example: while the reference design \citep[\emph{e.g.}][\citetalias{lsst}]{designSystem, designCamera} leads to a revisit time of 3 days on average for 
$18,000 ~\mathrm{deg}^2$ of sky, with two visits per night, this still allows for a large distribution and even a significant range of median values for the inter-night time gaps, as seen in  \autoref{fig:tgaps_example}.

{LSST will include several ``surveys,'' 
each helping to address the four key science pillars as well as other science goals in different ways.
The majority of the 10 years will be spent on a survey designed explicitly to meet the requirement specified in \cite{lsst}: the ``Wide-Fast-Deep'' survey (hereafter WFD). It is expected that this will take between 75\% and 85\% of the time on-sky. The remaining time will be spent on ``special programs'', including ``minisurveys'' (special coverage of extended areas of sky), ``Deep-Drilling-Fields'' (single pointings that will be visited periodically at an enhanced cadence and to reach a higher cumulative depth in the stacked images), and potentially ``Targets of Opportunity'' follow up of multi-messenger triggers.}

This loose division of LSST into the different flavors of (sub-) surveys 
implies different levels of flexibility for the observing strategies for different regions of the sky. For example, while the expected range of per-visit exposure times in the WFD region is  tightly constrained ($\sim30$ seconds) to achieve the goals of the four LSST science pillars and given observing efficiency constraints \citepalias{lsst}, some minisurveys may be better served by (or even require) different  exposure times, extending potentially much shorter and/or longer exposures than the WFD program exposure time.

{Perhaps uniquely among modern surveys, Rubin Observatory has embedded community involvement in the design of the survey \citep{frontpaper} and to that end has shared its extensive  simulations framework with the scientific community at a very high level of detail, including (but not limited to): detailed hardware specifications, facility operations models (including detailed observatory and instrument overheads), atmospheric transmission, and also models for astrophysical populations and interstellar dust which allow simulated recovery of tracer populations \citep{simsFramework}. For most users in the scientific community, it is the metadata of the predicted observing strategies (\emph{e.g.} observing time, expected seeing, instantaneous depth to $5\sigma$~photometric precision) that is most relevant to the evaluation of survey strategies: Rubin has made a large number (many hundreds to-date, see \citealt{frontpaper}) of simulated LSST surveys 
available to the community. The Operations Simulator, \citep{opsim}, generates the metadata for a full ten-year period of operation under specified desiderata for the run characteristics. 

Led by Lynne Jones \& Peter Yoachim at the University of Washington, the project has also developed a dedicated Metrics Analysis Framework (\maf: \citealt{maf})\footnote{Also available at \url{https://www.lsst.org/content/lsst-metrics-analysis-framework-maf}}, and continues to work with the community in the development of the tools to extract the scientific utility of the \opsim s for various scientific cases. Standard metrics run on all \opsim s by the project fall under the main \texttt{sims\_maf} package,\footnote{\url{https://github.com/lsst/sims_maf}} while community-contributed metrics are curated at the \texttt{maf-contrib} project.\footnote{\url{https://github.com/LSST-nonproject/sims_maf_contrib}}. We discuss the Operations Simulator and \maf~in a little more detail from the point of view of the true-novelties we seek, in  \autoref{ss:MAF}.} See also \citet{frontpaper} for more details.


Recent community input on the LSST survey strategy roughly divides into three phases. The first phase concluded with the development of the ``Community Observing Strategy Evaluation Paper'' (COSEP; \citealt{cosep}), which attempted to distill the requirements of a wide range of science cases into specifications (and in some cases evaluations) of simple quantitative measures of scientific effectiveness that could be compared between science cases. In the second phase,  the community was asked by the project to prepare cadence whitepapers to suggest alternatives to the baseline cadence; 46 whitepapers were ultimately submitted.\footnote{\url{https://www.lsst.org/submitted-whitepaper-2018}} 

In the current phase, the community and Rubin are working to implement the quantitative scoring for the various science cases to allow them to be compared on a timescale commensurate with the ultimate decisions by the project on the survey strategy to adopt. This paper forms part of this third phase of community input.


{A word on notation is in order.
{We refer to a simulated 10-year survey as an \opsim. }
Following the naming convention of the the \citetalias{cosep}, use the acronym ``\maf''(
or sometimes ``metric'') to refer to a piece of code that measures properties of an \opsim~on a per-field basis. The overall evaluation of a strategy requires assessing its power over a large number of scientific goals. Therefore, in order to be useful for comparison, the \maf s must be summarized themselves into Figures of Merit (\fom s): single numbers that convey the power of a survey (as simulated) to achieve a specific science goal. An example of a {\it metric} might be a characterization of the time-gaps between repeat observations of each field in a particular filter-pair of interest, while the associated \fom~would collapse the distribution into a single number that captures the sensitivity of the strategy to the detection of transients in some range of parameter space. For more on the operational definitions of metrics and \fom s, see the \citetalias{cosep}.}

As discussed in the introduction to the \citetalias{cosep}, ideally the \fom s would be measured in bits of information that the survey would contribute in excess of the previously-available information on a phenomenon. While clear in principle, this information-theory inspired definition of a \fom~is challenging to achieve in practice. Not all science cases easily translate into a measurement on a quantity. For cosmology, for example, one could conceivably quantify the scientific power of a survey by the decrease in the uncertainty on the scientific parameter of interest, for example $H_0$. However, the survey power even for identifying particular tracers becomes more ambiguous. The power of a survey in identifying progenitors of Supernovae, for example, is less easily quantifiable: as additional qualifiers are placed on the phenomena to be measured (for example: sensitivity to different types of progenitor), the translation of survey sensitivity into bits of additional information becomes increasingly difficult.  Following this logic, measuring the power of a survey to discover truly novel phenomena would be impossible.
The assessment of the ability of a survey realization (\opsim~run) to discover true novelties requires a model-free approach (otherwise we would by default limit ourselves to unobserved, but predicted, phenomena:  \citealt{AD2009}). 

We have set out to define metrics and \fom s that will allow comparison of simulated LSST strategies based on their potential to discover \emph{true novelties}, in terms of discovery parameter-space that is well-covered (or not!) by the simulated surveys. We make these \maf s and \fom s publicly available to aid Rubin survey strategy decisions.

The \opsim s are continually under development based on input from the project and the scientific community, and thus the suite of available simulations is continuously evolving. Improvements made between releases include general strategy updates (such as changes to the recommended exposure time per visit), improvements in the implementation of engineering constraints (such as the time required for filter changes) and improvements in the implementation of the observing strategies themselves (such as the way in which special cadences are implemented). \citet{frontpaper} provides more detail.\footnote{The release details are also announced on the LSST Community web forum, e.g. \url{https://community.lsst.org/t/survey-simulations-v1-7-1-release-april-2021/4865}.} 

We remain agnostic on the accuracy with which the \opsim s actually implement the desired strategies, but focus instead on the output: how well the resulting \opsim s support the detection of true anomalies as quantified in our metrics and figures of merit. We selected the \ovfive~family of simulations (a major release from 2020 May with 86 simulated strategies)\footnote{\url{https://community.lsst.org/t/fbs-1-5-release-may-update-bonus-fbs-1-5-release/4139}} to develop and demonstrate the metrics and figures of merit, as it contains sufficient variety among the simulations to elucidate the various requirements for detecting true anomalies.

As the simulations evolve, application of the figures of merit to more recent releases is then straightforward. As an example, we present the evaluation of our figures of merit to the \ovseven~(2021 January) and \ovsevenone~(2021 April) releases, which implement an updated exposure time per visit (2$\times$15s instead of the 1$\times$30s used in \ovfive).

This publication is one of a pair: in this communication (Paper I) we focus on detecting individual objects of interest in a multidimensional feature space that includes time coverage, filter coverage, star density, and total footprint on the sky. Inclusion of constraints from {\it proper motion}, which is rather more involved and also lends itself naturally to detection of previously-unknown {\it populations} and structures, is deferred to Ragosta et al. (Paper II). The present paper therefore does NOT directly address proper motion anomalies: the reader is referred to Paper II for those issues. 


This paper is organized as follows: \autoref{sec:method} summarizes the simulations and methods, and describes the feature space we use. Sections \ref{sec:timegaps}-\ref{sec:footprint} then communicate the metrics and figures of merit we have developed, and present the evaluations of the figures of merit over the wide-fast-deep main survey area, on a wide range of simulated observing strategies. Here we consider the following metrics: color and time evolution (\autoref{sec:timegaps}), integrated depth (\autoref{sec:depth}), and spatial footprint (\autoref{sec:footprint}). \autoref{sec:discussion} then applies the set of the figures of merit to the \opsim s chosen, first to the WFD region (\autoref{ss:mainsurvey}), then to the minisurveys (\autoref{ss:minisurveys}). The application of the figures of merit to the more recent \ovseven~set of simulations is presented in \autoref{ss:v1.7}.
In \autoref{sec:conclusion} we conclude with some recommendations on the usage and interpretation of the metrics and figures of merit we have developed. {Interactive tools we have developed to facilitate exploration of the multi-dimensional feature space are presented in the Appendix.}

\section{Methodology} \label{sec:method}

Here we summarize the simulations and methods used. \autoref{ss:MAF} summarizes the observations simulator and metrics analysis framework, both provided by the Rubin Observatory, in the context of our work. \autoref{ss:spatialselection} briefly discusses the tools by which we accomplish spatial selection, to isolate regions such as the Galactic Plane and Magellanic Clouds. The usage of {\it feature space} to identify discovery space for true-novelties is introduced in \autoref{ss:featurespace} and the output metadata produced by \opsim~in this context is summarized in \autoref{ss:opsim}.

\subsection{MAF and \opsim~}\label{ss:MAF}
The Operations-Simulator software\footnote{ \url{https://www.lsst.org/scientists/simulations/opsim}} \opsim~allows the generation of a simulated strategy based on a series of strategy requirements: for example total number of images per field per filter, including simulated weather, telescope downtimes, \emph{etc}. The input of an \opsim~run are the survey requirements (survey strategy) and the output is a database of observations with associated characteristics (\emph{e.g.} $5\sigma$ depth) which specify a sequence of simulated observations for the 10-year survey. The Rubin \opsim~went through several versions since its initial creation \citep{coffey} that primarily differ in the optimization scheme. 

The Metric Analysis Framework (\maf\footnote{\url{https://www.lsst.org/scientists/simulations/maf}}) API is a software package created by  Rubin Observatory \citep{maf} to facilitate the evaluation of simulated LSSTs to achieve specific science goals as measured by the strategy’s ability to obtain observations with specified characteristics. The \maf~interacts with  databases. The \maf~has been made public upon its creation to facilitate community input in the strategy design.
The \maf~enables selections of observations within an \opsim~primarily by \texttt{SQL} constraint which allows the user to select, for example, filters or time ranges (\eg~the first year of the survey). Further, the choice of \texttt{slicers} allows the user to group observations. For example, one may ``slice'' the survey by equal-area spatial regions, using the HEALPIX scheme of \citealt{healpix2005}. Throughout, we choose a \texttt{HealpixelSlicer} with resolution parameter \texttt{NSIDE}=16, corresponding to 

\subsection{Spatial selection}\label{ss:spatialselection}

In practice, \opsim~generates the synthetic observations using ``proposals'' for various assumed programs, including Deep Drilling Fields (DDFs), WFD, and ``special'' programs such as the Galactic mid-plane, Magellanic Clouds and the North Ecliptic Spur (NES, which affords greater sensitivity to Solar System objects; e.g. \citetalias{cosep}, Chapter 2), with the ``{\tt proposalID}'' parameters preserved in the \opsim~output. When evaluating our figures of merit for the WFD region, 
we use this ``{\tt proposalID}'' to select the relevant observations.


When evaluating our \fom s to the mini-survey regions (\autoref{ss:minisurveys}), we select entries spatially, as the science that can be extracted from observations of a particular spatial region depends only on what was observed, and not on the {\tt proposalID} with which each observation was originally identified. This is particularly relevant when considering simulated strategies that extend the WFD region to encompass regions that would be classified as ``minisurveys'' in the other simulated strategies: \citealt{olsenbigsky}, for example, discusses some possible strategies that would do this. The spatial selector we have developed is quite flexible - regions can be specified programmatically or by hand - and we have made it publicly available.\footnote{\url{https://github.com/xiaolng/healpixSelector}}

\subsection{Feature space}\label{ss:featurespace}

Anomaly detection is an important field of research with deep methodological ramifications \citep{AD2009, 
martinez20}. Notable advances in the field have been achieved in recent years across disciplines: from threat detection in defense and security~\citep[e.g.][]{sultani2018real}, to astrophysics~\citep{Soraisam20, Pruzhinskaya19, Ishida19, Aleo20,  vafaei2019flexible, martinez20, lochner2020astronomaly, doorenbos2020comparison} with the discovery of rare and possible unique astrophysical phenomena \citep[although we note that two of these ``true novelties'' were detected through crowd-sourced data analysis]{Voorwerp09, oumuamua, tabbysstar}.  
Anomaly detection is generally approached either through unsupervised or supervised learning learning techniques \citep[e.g.][]{10.1007/11881599_134, bishop2006, hastie2009elements}. In unsupervised learning, or \emph{clustering}, a  similarity metrics is defined in the available features space enabling the grouping together of similar objects together, as well as the identification of objects that do not belong to any existing group (the anomalous objects). Alternatively, the supervised approach identifies groups in a latent lower-dimensional space based on known classifications in the original feature space derived by domain experts (typically, deep learning approaches to anomaly detection belong to this category). 

Both of these approaches implicitly rely on the completeness of measurements in the original feature space: gaps in the observing strategy  affect the discovery of true novelties by both increasing the risk that an anomaly would go undetected, if it falls in a gap, and making its anomalous nature harder to assess. In this series of papers, we focus on survey design to maximize the throughput of algorithms for anomaly detection, regardless of the nature of the algorithmic approach.

As measured by imaging surveys, astronomical objects are characterized by  brightness, brightness ratio in different portions of the energy spectrum (color), position, shape, and the rate and direction of change in any of those features. The collection of properties defines 
a multidimensional phase space, with
different categories of phenomena lying in different regions of this phase space (see \autoref{fig:phasespace}). Accordingly, we identified the following features that can be measured in the Rubin Observatory data:
\begin{itemize}
\item Color
\item Time evolution
\item Motion    
\item Morphology
\item Association
\end{itemize}

We set morphology aside, as largely the power of the survey to measure morphological anomalies does not depend on the survey strategy, but rather on the image system design (\eg,  resolution and depth). We assume that measuring anomalous associations depends on our accuracy in measuring the properties of each object.

{To measure dynamical anomalies in a completely model-independent way proves to be more involved, because it requires comparison of measured proper motions to those of established Galactic dynamical parameters. {\it Motion} is thus deferred to Paper II, where we also develop a Figure of Merit for the detection of previously unknown {\it populations.}}

Having identified features that can be extracted from the Rubin Observatory data, and the LSST in particular, such as color information or lightcurve evolution, we measure the completeness of the survey in a hypercube in the feature space as a model-independent measure of the power to detect \emph{novel  transients} or novel modes of variability.   Generally, we define transients as objects whose observational \emph{and physical} properties are changed by some event, usually as the result of some kind of eruption, explosion or collision, whereas variables are objects whose nature is not altered significantly by the event (\eg, flaring stars). Furthermore, some objects vary not because they are intrinsically variable, but because some aspect of their viewing geometry causes them to vary (\eg, eclipsing binaries). 

One further parameter that influences our ability to detect anomalies is the sky footprint. Trivially, a larger sky footprint will lead to a higher event rate for anomalies. If one wants to maximise the chance of detecting extragalactic anomalies then a larger footprint would be favorable, while the probability of detecting galactic anomalies will scale with density of objects in the sky. And both will scale with the depth over which the footprint is observed.


Ultimately, we define a set of metrics that can simply be summed to generate a \fom~for \emph{true novelties}:

\begin{equation}
   FoM = \sum_{i={c,s,d,
   A_\mathrm{sky}, D_\mathrm{Star} }} w_i ~\fom_i
   \label{eq:fom:master}
\end{equation}
where $c,~s,~d,~A_\mathrm{sky},~D_\mathrm{Star},$ represent the color, lightcurve shape, magnitude depth, footprint, and star density respectively, and $w$ are weights that can be assigned to favor the discovery of, for example, transients over non-evolving objects, or galactic over extragalactic transients. 


{The weights $w_i$~allow the investigator to imprint their own judgement on the relative scientific importance of the different metrics. Because we wish to remain as phenomenon-agnostic as possible,} we refrain from assigning weights. Instead, we normalize each \maf~to the best of our ability in a 0-1 range where 1 is optimal,
so as to provide a ``neutral'' comparison of the existing LSST simulations.

\subsection{\opsim~Data}\label{ss:opsim}
We base our results primarily on \ovfive,  a recent \opsim ~run that contains 86 databases in 20 families as listed in \autoref{tab:opsim}. 

\startlongtable
\begin{deluxetable}{clc}
\tabletypesize{\footnotesize}
\tablewidth{0pt}

\tablecaption{ \ovfive~ databases \label{tab:opsim}}
\tablehead{\colhead{ Family} & \colhead{Name}  }
\startdata 
 agn &agnddf  \\ 
 \hline 
 \multirow{2}{*}{alt} & alt\_dust \\
 & alt\_roll\_mod2\_dust\_sdf\_0.20  \\ 
 \hline
 \multirow{3}{*}{baseline } 
 & baseline\_2snaps  \\ 
 & baseline\_samefilt \\
 & baseline   \\ 
 \hline
 \multirow{6}{*}{bulges}  &bulges\_bs  \\ 
 &bulges\_bulge\_wfd  \\ 
 &bulges\_cadence\_bs    \\ 
 &bulges\_cadence\_bulge\_wfd    \\ 
 &bulges\_cadence\_i\_heavy    \\ 
 &bulges\_i\_heavy    \\ 
 \hline
 daily & daily\_ddf \\
 \hline
 \multirow{6}{*}{dcr} 
 &dcr\_nham1\_ug \\ 
 &dcr\_nham1\_ugr  \\ 
 &dcr\_nham1\_ugri  \\ 
 &dcr\_nham2\_ug   \\ 
 &dcr\_nham2\_ugr  \\ 
 &dcr\_nham2\_ugri \\
 \hline
 descddf &descddf    \\ 
 \hline
 \multirow{8}{*}{filterdist} 
 &filterdist\_indx1    \\ 
 & filterdist\_indx2 \\
 & filterdist\_indx3 \\
  & filterdist\_indx4 \\
 & filterdist\_indx5 \\
 & filterdist\_indx6 \\
 & filterdist\_indx7 \\
 & filterdist\_indx8 \\
 \hline 
 \multirow{12}{*}{footprint}&footprint\_add\_mag\_clouds   \\ 
 &footprint\_big\_sky\_dust    \\ 
 &footprint\_big\_sky\_nouiy    \\ 
 &footprint\_big\_sky    \\ 
 &footprint\_big\_wfd \\
 &footprint\_bluer\_footprint    \\ 
 &footprint\_gp\_smooth    \\ 
 &footprint\_newA    \\ 
 &footprint\_newB    \\ 
 &footprint\_no\_gp\_north    \\ 
 &footprint\_standard\_goals    \\ 
 &footprint\_stuck\_rolling    \\ 
 \hline
 \multirow{5}{*}{goodseeing} 
 &goodseeing\_gi \\ 
 &goodseeing\_gri    \\ 
 &goodseeing\_griz    \\ 
 &goodseeing\_gz   \\ 
 &goodseeing\_i   \\ 
 \hline 
 greedy & greedy\_footprint \\
 \hline
 roll & roll\_mod2\_dust\_sdf\_0.20   \\ 
\hline
\multirow{6}{*}{rolling} 
 &rolling\_mod2\_sdf\_0.10    \\ 
 &rolling\_mod2\_sdf\_0.20    \\ 
 &rolling\_mod3\_sdf\_0.10    \\ 
 &rolling\_mod3\_sdf\_0.20    \\ 
 &rolling\_mod6\_sdf\_0.10    \\ 
 &rolling\_mod6\_sdf\_0.20    \\ 
 \hline 
 \multirow{4}{*}{short} 
 &short\_exp\_2ns\_1expt    \\ 
 &short\_exp\_2ns\_5expt    \\ 
 &short\_exp\_5ns\_1expt    \\ 
 &short\_exp\_5ns\_5expt    \\ 
 \hline
 spider &spiders    \\ 
 \hline
 \multirow{6}{*}{third}
 & third\_obs\_pt120  \\ 
 & third\_obs\_pt15  \\ 
 & third\_obs\_pt30  \\ 
 & third\_obs\_pt45  \\ 
 & third\_obs\_pt60  \\ 
 & third\_obs\_pt90 \\
 \hline
 \multirow{4}{*}{twilight neo}&twilight\_neo\_mod1    \\ 
 &twilight\_neo\_mod2    \\ 
 &twilight\_neo\_mod3    \\ 
 &twilight\_neo\_mod4    \\ 
 \hline
 u60&u60    \\ 
 \hline
 var & var\_expt \\
 \hline
 \multirow{16}{*}{wfd}
 &wfd\_depth\_scale0.65\_noddf    \\ 
 &wfd\_depth\_scale0.65    \\ 
 &wfd\_depth\_scale0.70\_noddf    \\ 
 &wfd\_depth\_scale0.70    \\ 
 &wfd\_depth\_scale0.75\_noddf    \\ 
 &wfd\_depth\_scale0.75    \\ 
 &wfd\_depth\_scale0.80\_noddf    \\ 
 &wfd\_depth\_scale0.80    \\ 
 &wfd\_depth\_scale0.85\_noddf    \\ 
 &wfd\_depth\_scale0.85    \\ 
 &wfd\_depth\_scale0.90\_noddf    \\ 
 &wfd\_depth\_scale0.90    \\ 
 &wfd\_depth\_scale0.95\_noddf    \\ 
 &wfd\_depth\_scale0.95    \\ 
 &wfd\_depth\_scale0.99\_noddf \\ 
 &wfd\_depth\_scale0.99  \\
\enddata
\tablecomments{The description of these \opsim s can be found in the release note of \ovfive}
\footnote{\url{https://community.lsst.org/t/fbs-1-5-release-may-update-bonus-fbs-1-5-release/4139} }
\end{deluxetable}
A more detailed description and discussion of the simulations can be found in \citet{frontpaper} and on the Community LSST discussion forum.\footnote{\url{https://community.lsst.org/t/fbs-1-5-release-may-update-bonus-fbs-1-5-release/4139} \url{fbs-1-5-release-may-update-bonus-fbs-1-5-release}, released in May 2020}
Here we simply note that \texttt{baseline} refers to the straightforward implementation of the requirements in \citet{lsstSRD}; the acronyms that were mentioned in this work to refer to different surveys within LSST, such as \emph{WFD} (Wide Fast Deep) and \emph{DDF} (Deep Drilling Fields) are mirrored in the names of the families of \opsim s. We note that \emph{roll} refers to ``rolling cadence'': a WFD strategy implementation where fields are not observed homogeneously in time over the survey lifetime, but rather some fields are observed more frequently early in the survey and, to different degrees, abandoned later on, to focus on other fields. These strategies provide a denser cadence on each field for some fraction of the survey time, while preserving the overall cumulative depth requirements, and are generally beneficial to the study of  rapid-time-scale transients (including supernovae). The \texttt{footprint} family of observations modifies the survey footprint according to different recommendations.\footnote{\url{https://www.lsst.org/call-whitepaper-2018}} \texttt{Pair strategy} is a family of \opsim s that explores different approaches to pairing in time filters and observations. The \texttt{filterdist} varies the filter distribution across WFD and \texttt{third} adds a third observation at the end of the night. The \texttt{goodseeing} family explores different requirements on weather to enable observations. The remaining \opsim~families explore exposure time (\eg, \texttt{short} or \texttt{var expt}), specific observing times (exclusively or in combination with the regular surveys) such as twilight, explicit observing phenomena that can be enhanced by cadence choices such as \texttt{dcr}, differential cromatic diffraction, or synergy with other surveys, such as Euclid.
Lastly, \texttt{u60} has longer exposures in the  $u$ band (60 second compared to the standard 30 seconds) and some \opsim s explore modifications of the single exposure time by implementing a single 30 seconds observation instead of 2x15 ``snaps'' (that get combined into a single image to produce the standard Rubin data products, see also \autoref{ss:v1.7} for an assessment of the impact of this choice across \opsim s). 

\section{Color and time evolution}\label{sec:timegaps}
\begin{figure*}[t!]
\centering 
\includegraphics[scale=0.35]{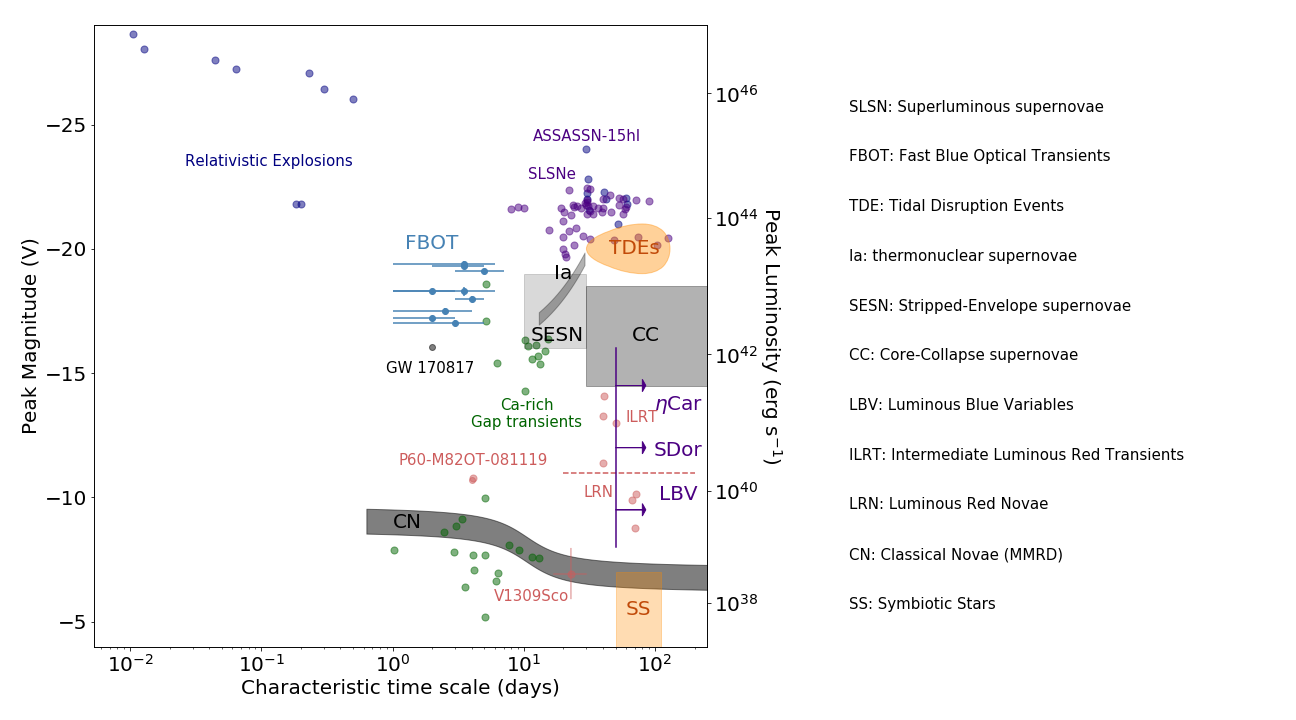}
\caption{The phase space of transients, reproduced with 
permission from \citet{lsst} with minor modification: the intrinsic brightness is plotted against the characteristic time scale of evolution. Shaded areas indicate the  region of this phase-space occupied by various classes of objects, with individual objects indicated for some of the the less populated classes. The notable gap at all intrinsic magnitudes fainter than -20 is likely due, at least in part, to an observational bias, as surveys are typically not able to probe large volumes of the Universe at short time-scales down to faint apparent magnitudes. Data for Superluminous supernovae (SLSNe) and Fast Blue Optical Transients (FBOTs) that were not included in the original plot are collected from \citealt{Drout14,Inserra19} respectively. See \autoref{ss:featurespace}.}
\label{fig:phasespace}
\end{figure*}

\begin{figure*}[t!]
\centering
\includegraphics[scale=0.3]{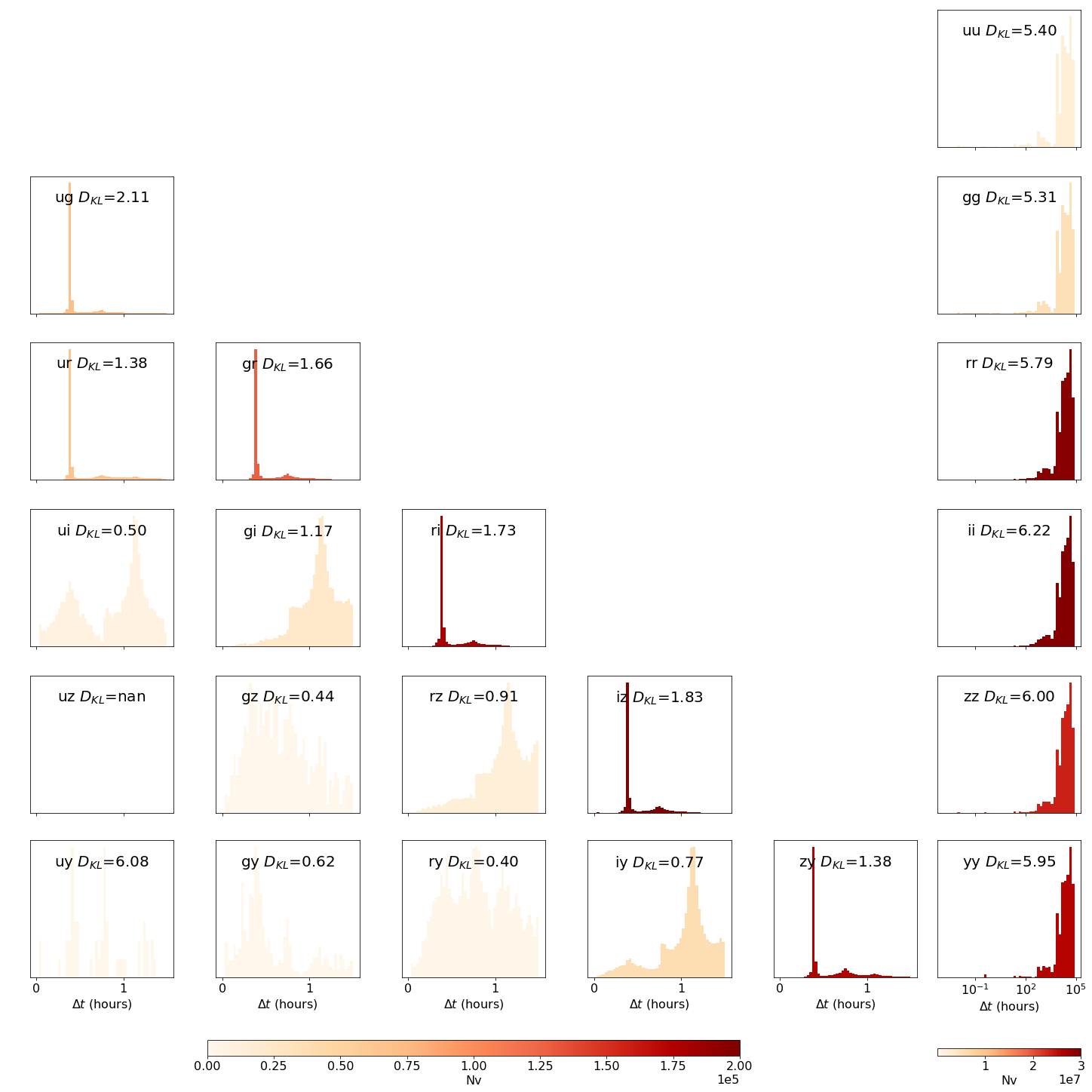}
\caption{
The distribution of all time gaps for the  \texttt{baseline\_v1.5} \opsim. The triangle of plot on the left shows all time gaps between different filters (which enable the measurement of color) within 1.5 hours. The column of plots on the right shows the distribution of time gaps in the same filters for the 10-year survey, which enables the measurement of brightness changes. The filters are indicated in each quadrant: from $u$ to $y$ moving from top to bottom and left to right. All histograms are normalized but the intensity of the color is proportional to the to total number of observations in that filter-pair, as indicated by the color bar. In each quadrant the value of $D_{KL}$ is  reported (see \autoref{ss:tgaps}). We note that the majority of observations are taken with adjacent filters, which gives a narrow leverage on the spectral energy distribution (SED), and less power to measure color. Color is in fact better measured with filters that are more separated in wavelength, for example \emph{g-i} or \emph{r-z}, as described in \citet{Bianco_2019}}. 
\label{fig:tGaps_dist}
\end{figure*}

\begin{figure*}[t!]
\centering
\includegraphics[scale=0.5]{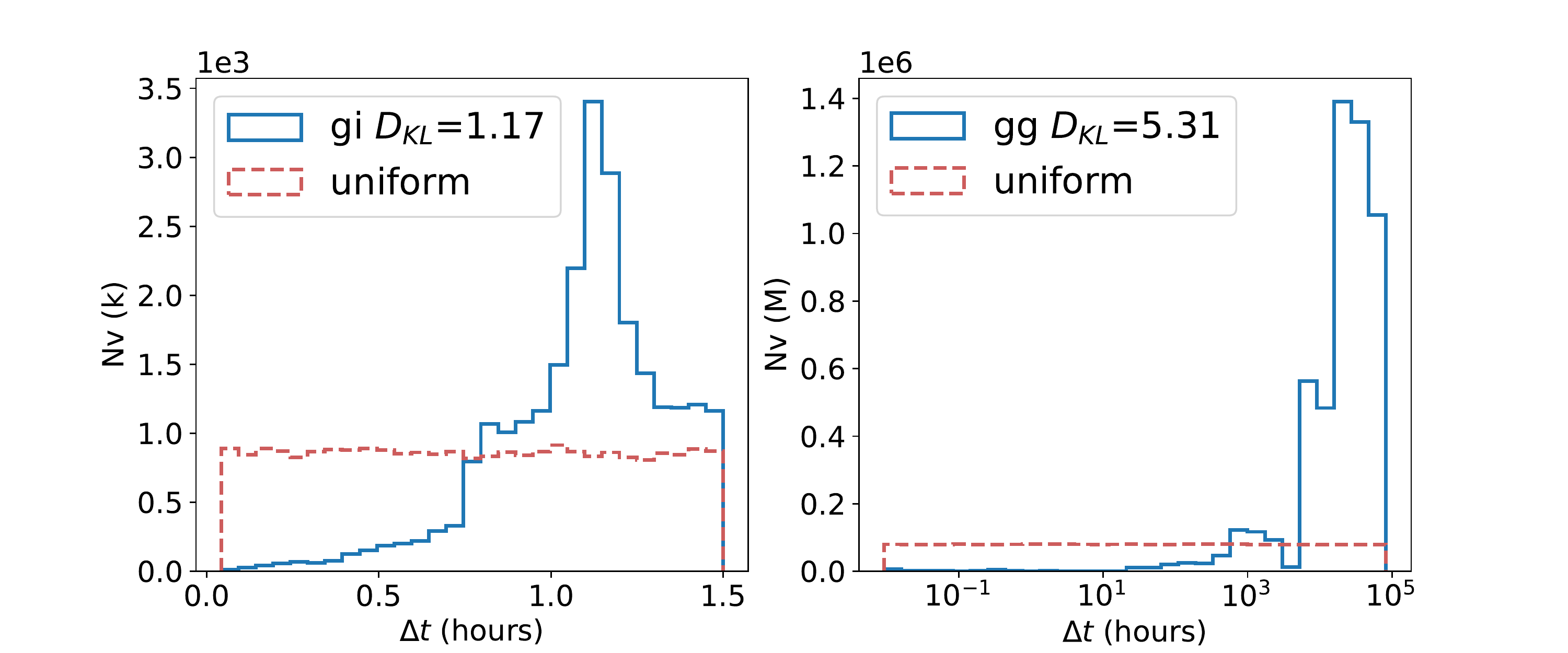}
\caption{
Time gaps in $g-i$ (left) and $g-g$ (right) from \texttt{baseline\_v1.5} compared to the ``ideal'' distribution, plotted in orange: a uniform distribution for the colors (left, with the $y$-axis in units of 1000 observations) and a uniform distribution in log space for the lightcurve shape (right, with the $y$-axis in units of 1 million observations). See \autoref{ss:tgaps}. 
}
\label{fig:tGaps_gi}
\end{figure*}
 
 \begin{figure*}[!th]
 \centering
\gridline{
  \fig{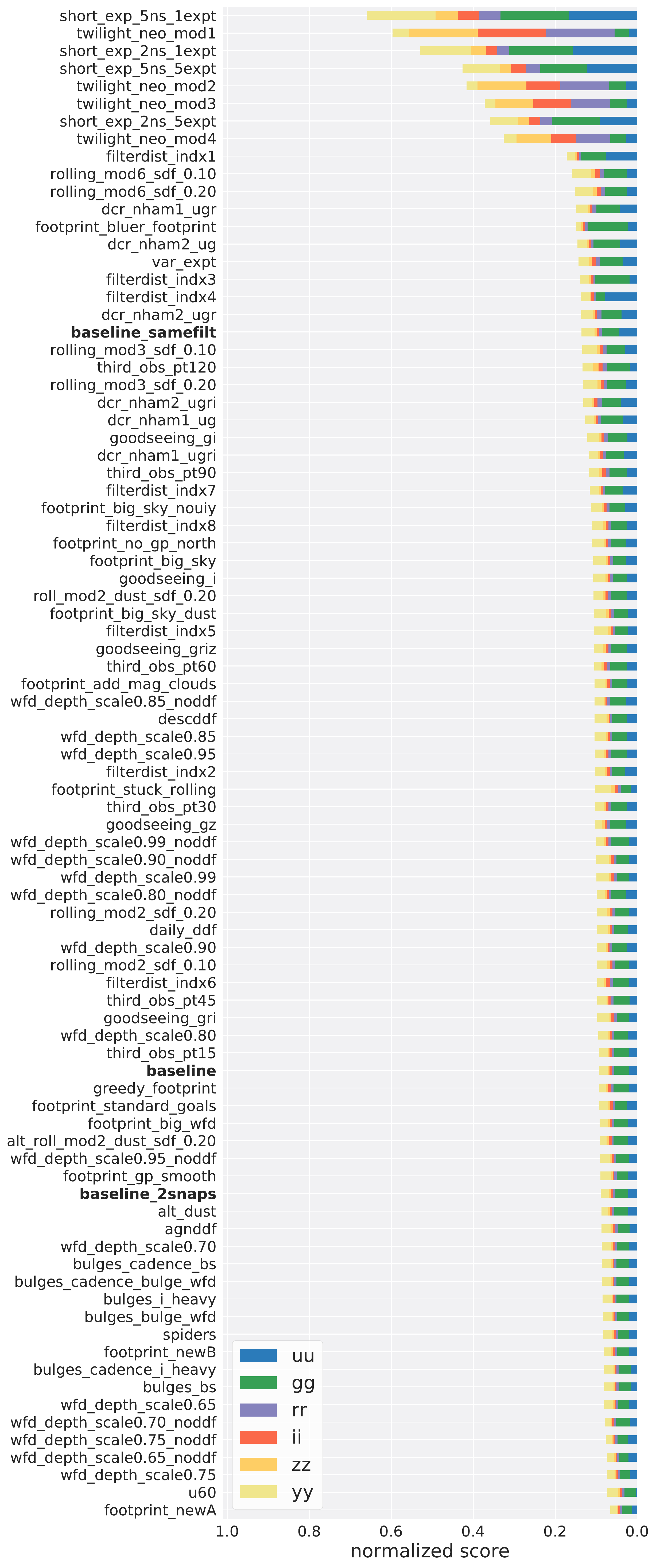}{0.4\textwidth}{($a$)} 
  \fig{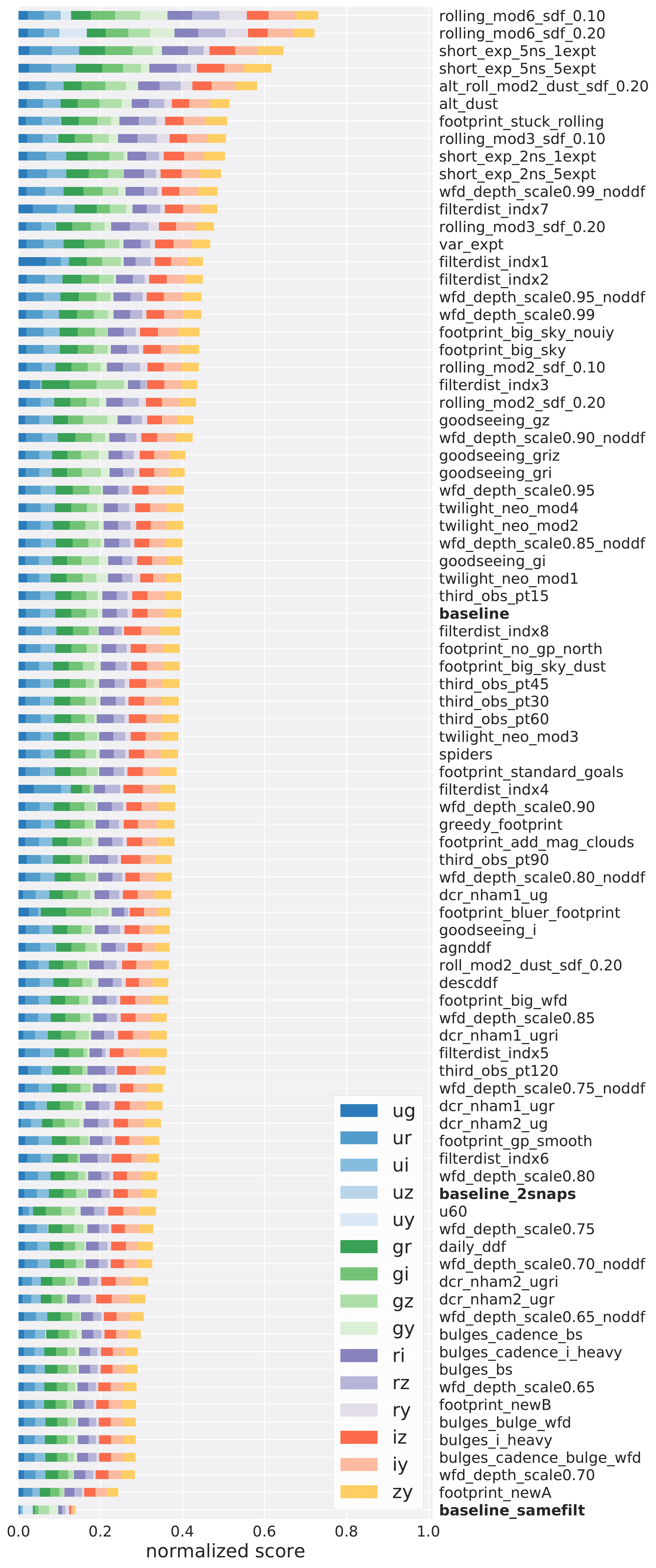}{0.4\textwidth}{($b$)}
 }

\caption{Figures of merit \fom$\mathrm{tGaps}$ for the \ovfive~runs based on the distribution of time gaps. The \fom$\mathrm{tGaps}$  is calculated as described in  \autoref{eq:fom:tgaps}. The plot on the \emph{left} ($a$) shows the \fom~ for repeat visits in the same filter. The plot on the \emph{right} ($b$) shows the value for observations in pairs of different filters. Each \opsim~ is presented as a bar whose length corresponds to the value of the \fom: the \fom s for different filters are concatenated horizontally. For example: on the left the different color bars represent the time gap \fom~ for different filters from $u$ to $y$. The \opsim s are sorted by the total \fom. In subplot ($a$) the bars grow toward the left, in ($b$) toward the right, so that asymmetries in the plot can give intuition on the overall distribution of the two different metrics across the set of the \opsim s.
 See  \autoref{ss:tgaps}.  
}
\label{fig:barh_tgaps_wfd}
\end{figure*}

Astrophysical transients and variable phenomena captured humanity's curiosity through the history of science. Modern astrophysics and particularly the use of digital equipment in the last half-century enabled extremely fast paced advances in this field.  \autoref{fig:phasespace}, reproduced and updated from \citet{lsst}, shows the phase space of known astrophysical transients: transients and variable phenomena occupy different regions of this phase space of intrinsic brightness \emph{vs} characteristic time scales. At the beginning of the 20th century, essentially only supernovae were known to exist, and the phase space populated rapidly with many different classes of transients since. It is worth noting the gap for timescales shorter than $\sim 1~\mathrm{day}$: while it is possible that this region is scarcely populated {\it intrinsically}, it is also true that an observational bias impairs discovery in this region:  to be effective in discovery and characterization at these time scales surveys  need to reach high depth and high cadence simultaneously, while also surveying a large volume if phenomena in this region of the phase space are truly rare.

Due to their diversity in time scales, color, and evolution, the study of transients and particularly studies that aspire to discover new transient phenomena, requires dense space \emph{and} time coverage.  
The LSST has both high photometric sensitivity and a large footprint, enabling the surveying of a large volume of Universe. This offers tremendous opportunities to study the variable sky. 
LSST's capability to discover novel transients then largely depend on its observation cadence.

Different transients will benefit from different observation strategies because of the different phenomenological expression of their intrinsic physics. To make sure the observation strategies under design maximize  our chances to discover \emph{any} novel transient, we created the {\it filterTGapsMetric}. This \maf~evaluates  the ability of LSST's observation strategies to capture information about color and its time evolution at multiple time scales. We know some timescales remain unexplored in the present collection of LSST simulations, as discussed for example in the work of \citealt{bellm21} and \citealt{Bianco_2019}.

\subsection{The \texttt{filterTGapsMetric}} \label{ss:tgaps}

Rubin LSST will image the sky in six filter bands {\it u, g, r, i, z, y}. The \texttt{ filterTGapsMetric} measures all time gaps between two filters in an \opsim, \ie, {\it ug, gr, ri} and so on. The \texttt{filterTGapsMetric} \fom~evaluates the coverage of time gaps for each filter-pair. 


On a field-by-field basis, for each filter pair, the metric and \fom~ are evaluated as follows:

\begin{itemize}
    \item select the survey (\emph{e.g.} WFD in this paper) and the observation time range using \texttt {SQL} constraint and slice the sky with HealpixelSlicer (see \autoref{ss:MAF}); 
    \item fetch observation times for each field for all visit in either of the two filters;
    \item compute all possible time gaps {that can be constructed from pairs of} visits;
\end{itemize}
\autoref{fig:tGaps_dist} shows the distribution of time gaps for all filters pairs for the \texttt{baseline v1.5}. 

Armed with field-by-field time-gap distributions, the \fom~ for the entire candidate survey strategy is then computed by measuring how well the distribution of time gaps matches an ideal distribution. 
We use the Kullback-Leibler (KL) divergence~\citep[or relative entropy][]{kullback1951} to measure the discrepancy between the ideal and observed distribution. The KL divergence provides an information-criteria based measure of the difference between two  distributions: the KL divergence from $Q$ to $P$ is defined as ${D_{KL}(P||Q) = \sum P ~{\rm log}(\frac{P}{Q})} $. The KL divergence is not a distance (in the sense that it does not satisfy the triangle inequality), is in general not symmetric {(under exchange of $Q$~and $P$)}, and it is not normalized. To derive a normalized quantity from ${D_{KL}}$ we use ${ e^{-D_{KL}} }$, where two identical distributions, with  ${D_{KL} = 0}$  would contribute 1 to the sum, while ${D_{KL} > 0}$  would contribute $<1$. Thus a larger \fom~ would indicate a lower discrepancy from the ``ideal'' distribution and thus a scientifically preferable simulation. This \fom~ is naturally normalized between 0 and 1 for each field. 

{All that remains is to choose the ``ideal'' distribution of time-gaps against which candidate strategies will be compared. We choose different ``ideal'' distributions depending on whether color evolution or the lightcurve shape is being probed (\autoref{fig:tGaps_gi} shows an example).}
\citealt{Bianco_2019} have shown that color can be measured reliably even for rapid explosive transients, {for time-gaps as long as} 1.5 hours. Of course this is not necessarily true for novel phenomena, but we will use this as a fiducial time interval and take the ideal distribution to be a uniform distribution between 0 and 1.5 hours.
To probe lightcurve shapes {via pairs of observations in the same filter}, we want to measure evolution at all time scales, ideally down to the minimum possible repeat time of a few seconds set by the shutter and readout electronics. For observation-pairs in the same filter, then, we adopt a uniform distribution in
${\log_{\rm 10}(\Delta t)}$ for the entire 10-year survey.

The steps of the calculation of the \fom~for time gaps, are thus:
\begin{itemize}
    \item Compute the discrepancy-measure ${e^{-D_{KL}}}$ between the distribution of time gaps and an ``ideal'' distribution, for each filter-pair. This step is shown in \autoref{fig:tGaps_dist}.
    \item Sum the discrepancy-measures over the filter-pairs, weighted by the number of visit-pairs over the whole sky in each filter-pair ${N_k}$, and optionally by a ``scientific'' weight-factor $w$~that allows certain filter-pairs {and/or spatial fields} to be (de)-emphasized. 
\end{itemize}
This weighted sum, {over the filter-pairs and over the positions in the sky,} is the \fom~ for the \opsim of interest.
The process is summarized in the relation:
\begin{equation}
    \fom_{\mathrm{tGaps}} =\sum_i^{\mathrm{fields}} \sum_{k=ug, gr,...}^{\mathrm{pairs}} w_{k,i} {N_{k,i}} e^{-D_{KL,k,i}},
    \label{eq:fom:tgaps_i}
\end{equation}
where ${0 \le w_{k,i} \le 1.0}$ , ${N_k}$ stands for the number of visits in the \opsim~{for each of the filter-pairs}, and the index $i$ runs through the \texttt{healpixels}  \autoref{ss:MAF}). 

In practice, we use a simplified version of the above relationship where the {metrics are summed over the sky for each filter-pair before computing the} KL divergence,
since we will embed preferences in the pointing with subsequent components of the \fom (see \autoref{sec:footprint}):

\begin{equation}
    \fom_\mathrm{tGaps} = \sum_{k=ug, gr,...}^{\mathrm{pairs}} w_{k} {N_{k}} e^{-D_{KL,k}}
    \label{eq:fom:tgaps}
\end{equation}

As indicated earlier, we do not choose any weights: the value of $w_{k}$ is always set to 1 in our calculations. Some filters and filter combinations may well be more useful than others to discover anomalies. Trivially, the value of $w_{k}$ could be set by the limiting magnitude for the shallowest filter in a filter pair. 

\subsection{Results}\label{ss:tgapsresults}

 \autoref{fig:barh_tgaps_wfd} shows the \fom$_\mathrm{tGaps}$ calculated in  \autoref{eq:fom:tgaps} for all \opsim~ runs in \opsim~v1.5. {Because the ``ideal'' comparison distribution is different for the color (different-filter) and lightcurve shape (same-filter) pairs, expression (\ref{eq:fom:tgaps}) is evaluated twice for each \opsim: once over the 15 different-filter pairs (for color) and once over the six same-filter pairs, with the results presented separately.}
 
 Two families of \opsim s rise to the top of the list when ranked by \fom$_\mathrm{tGaps}$~  
 for the color diagnostics: \texttt{short} and \texttt{twilight} (panel $a$~of \autoref{fig:barh_tgaps_wfd}). This can be explained by the fact that these \opsim s~contain short exposures that fill in the distributions at short time scale. 
After a significant performance step we the see \texttt{filterdist, rolling, and dcr} families as the next best options.
 
The light curve shape
 \fom$_\mathrm{tGaps}$~(panel $b$~of \autoref{fig:barh_tgaps_wfd}),  shows the \texttt{rolling} family of \opsim s rising among the top performers: a rolling strategy naturally provides a log-like coverage which supports the discovery and study of transients at multiple scales. All top 10 performing surveys, from the point of view of the lightcurve-shape characterization, are \texttt{rolling} or \texttt{short} cadence \opsim s although we see a smooth performance decline with no sharp transition.


\section{Depth metrics}\label{sec:depth}

Because our time-gaps metrics are essentially based on the number of images that meet some criteria in an \opsim, it is important to assure that the images that are counted are all meeting some quality standards. 
In particular, we need to include information about the image depth (\ie~limiting magnitude), so that we compare the discovery potential within the same volume of the Universe. Some \opsim s augment the the WFD survey with short exposures (see \autoref{ss:opsim}). In fact, we noted in the $FoM_\mathrm{tGaps}$ analysis (\autoref{sec:timegaps}) that \opsim s that include short exposures rise to the top of the ranked list of \opsim s: while these \opsim s meet the nominal criteria and provide valuable image pairs at short time gaps, they may fail to extend the survey volume to unexplored regions,
which is the most important contribution LSST will make in the anomaly discovery space.  To account for this, we add a metric component that measures the depth of the images collected by an \opsim. 

We inspect the depth distribution of the \opsim s for each filter, comparing them to the apparent magnitude limits specified in the Science Requirements Document \citep[Table 6 in][]{lsstSRD}. 
{In practice, the main contributor to the difference in limiting depth between the \opsim s seems to be the time allocated to short exposures.} Short exposures are typically designed for specific purposes, such as the detection of Near Earth Objects (NEOs) (\eg, the \texttt{twilight\_neo} family) or decreasing the saturation limit so as to enable calibrations with shallower surveys \citep{gizis19}. We want to penalize surveys where these short exposure come at a cost of deeper images. 

\begin{figure}
    \centering
    \includegraphics[scale=0.4]{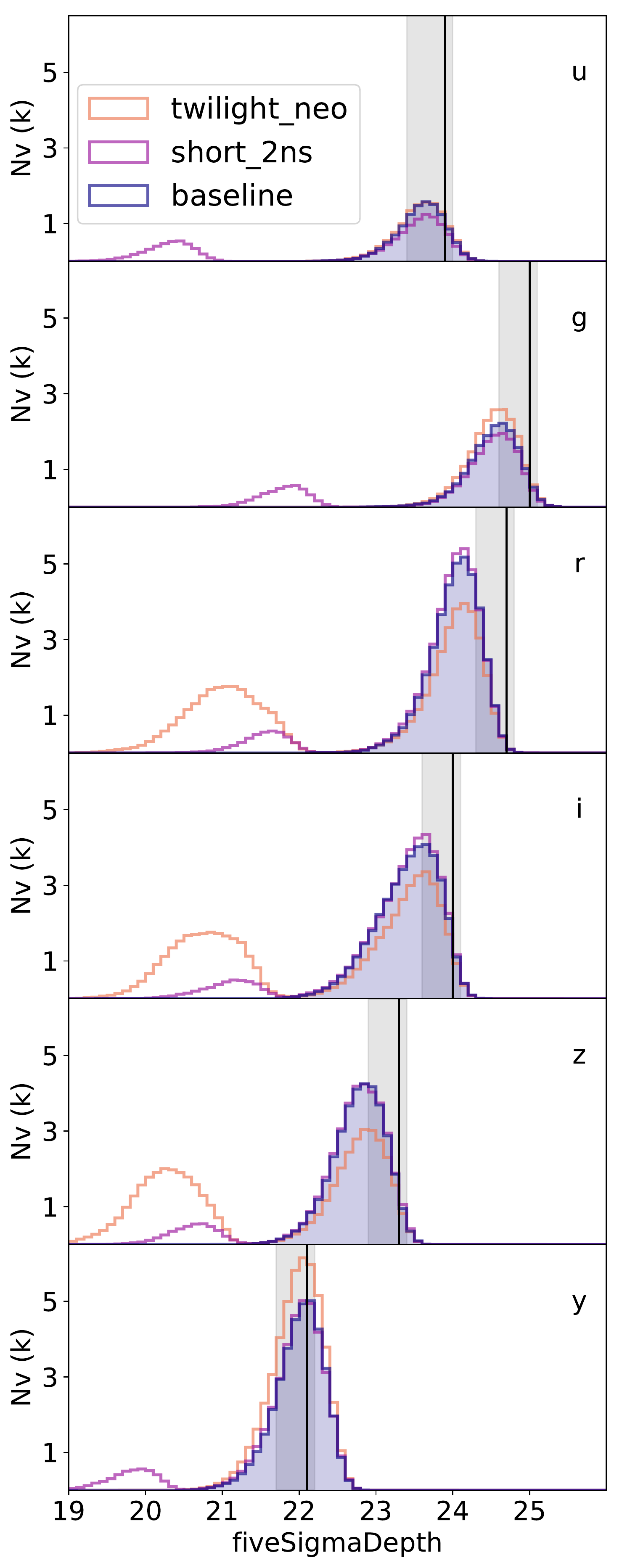}
    
    \caption{Distribution of depth for images in three different \opsim s. The survey specifications are indicated as a gray band (minimum requirement to stretch goal) and a vertical line (design specification) as per \cite{lsstSRD} table 6 for each filter (as indicated in the top left of each panel). Some \opsim s are designed to include in the WFD short exposures and may perform well in metrics based on number of exposures taken. But shallower images generate lower SNR measurements and only allow the exploration of a smaller volume of the Universe: this would have negative impact on the discovery of anomalies if it came at the cost of long exposures. We show the distribution of 5-$\sigma$ depth for images in three \opsim s: the baseline (blue filled histogram), \texttt{twilight\_neo\_mod1}, and \texttt{short\_exp\_2ns\_1expt}. \texttt{twilight\_neo\_mod1} and \texttt{short\_exp\_2ns\_1expt}  have additional short-exposures leading to a bimodal  distribution. However, while for  \texttt{short\_exp\_2ns\_1expt} \opsim~ the distribution of faint (fainter than $\sim21.5$ in each band) images is similar to the baseline's one, \texttt{twilight\_neo\_mod1} has fewer faint images in $r$, $i$, and $z$ band, and more in $y$ band. See \autoref{sec:depth}.}
    \label{fig:depthhist}
\end{figure}

\autoref{fig:depthhist} compares the per-image limiting magnitude (at $5\sigma$) distribution for two implementations of short-exposures (\texttt{twilight\_neo\_mod1} and \texttt{short\_exp\_2ns\_1expt}) with the baseline survey (blue filled histogram).  
The \opsim s including short exposures show a
 bimodal distribution of limiting magnitude. The short exposures contribute to a cluster that peaks at magnitude brighter than 21 in any bands ($u$=20.45, $g$=20.95, $r$=20.95, $i$=20.95, $z$=20.75, $y$=19.95 for \texttt{short\_2ns} and $r$=20.95, $i$=20.85, $z$=20.25, $y$=20.95 for \texttt{twilight\_neo\_mod1}). However, while for \texttt{short\_exp\_2ns\_1expt} the distribution of faint (fainter than magnitude $\sim21.5$) images is not substantially different from that of the baseline survey, \texttt{twilight\_neo\_mod1} has fewer faint images in $r$, $i$, and $z$ band, and more in $y$ band. 
    
The related \fom~is then the difference between the median of the distribution and the survey specification in the Science Requirements Document \citep{lsstSRD}:

\begin{equation}
\fom_\mathrm{ depth} = \frac{1}{6}\sum_{k=u, g, r, i, z, y}^{\mathrm{filters}} (m_{\mathrm{median},~k} - m_{\mathrm{goal},~k})n_k,
\label{eq:depth}
\end{equation}
where the sum extends to the six filters and $n_k$~is a numerical factor that scales the range of the \fom~for each filter to $[0, 1]$. This has the effect of treating the contributions from each filter equally.
An \opsim~must therefore rank at the top in all filters simultaneously to achieve \fom$_\mathrm{depth}$=1.0.

\begin{figure}
    \centering
    \includegraphics[scale=0.3]{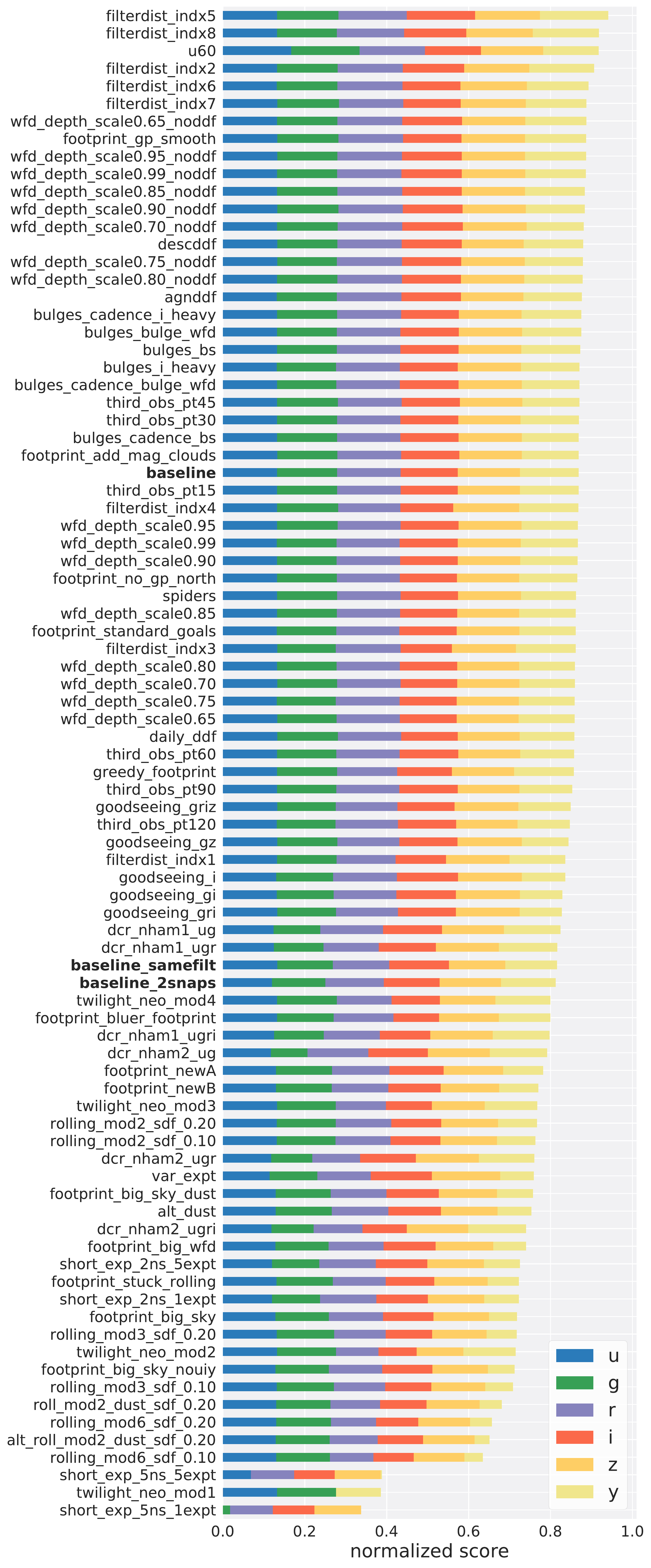}
    \caption{Ranking of \opsim s based on the depth of the exposure as discussed in \autoref{sec:depth}. \texttt{u60}, which produces 60-second $u$-band exposure instead of the standard 30-second, extends the observed volume slightly in $u$ band and ranks highly in this metric, but was performing poorly in both \fom$_\mathrm{tGaps}$ (\autoref{fig:barh_tgaps_wfd}, a and b). Otherwise, for the most part, family of \opsim s are clustered together in this diagram, all with similar $FoM_{depth}$ score: 90\% of the \opsim s generate values with 10\% of each other in this metric.}
    \label{fig:barh_depth}
\end{figure}

This leads to a ranking of the \opsim s shown in \autoref{fig:barh_depth}, the short exposure family of images ranks low, compensating for the high rank conferred in the earlier \fom~by the higher number of images. Aside from the \texttt{u60, short}, and \texttt{twilight} families, all other \opsim s have a similar score, between $\sim0.8$ and $\sim0.9$. The \texttt{u60}, which produces 60-second $u$-band exposure instead of the standard 30-second, ranks near the top.  We also note that the \texttt{rolling} and \texttt{footprint\_big\_sky} families are penalized in this metric. This may be a consequence of the added constraints on pointing competing with the constraints on image quality (which relate to weather, airmass, etc.).

\section{Footprint}\label{sec:footprint}
\begin{figure*}
\centering
\includegraphics[scale=0.3]{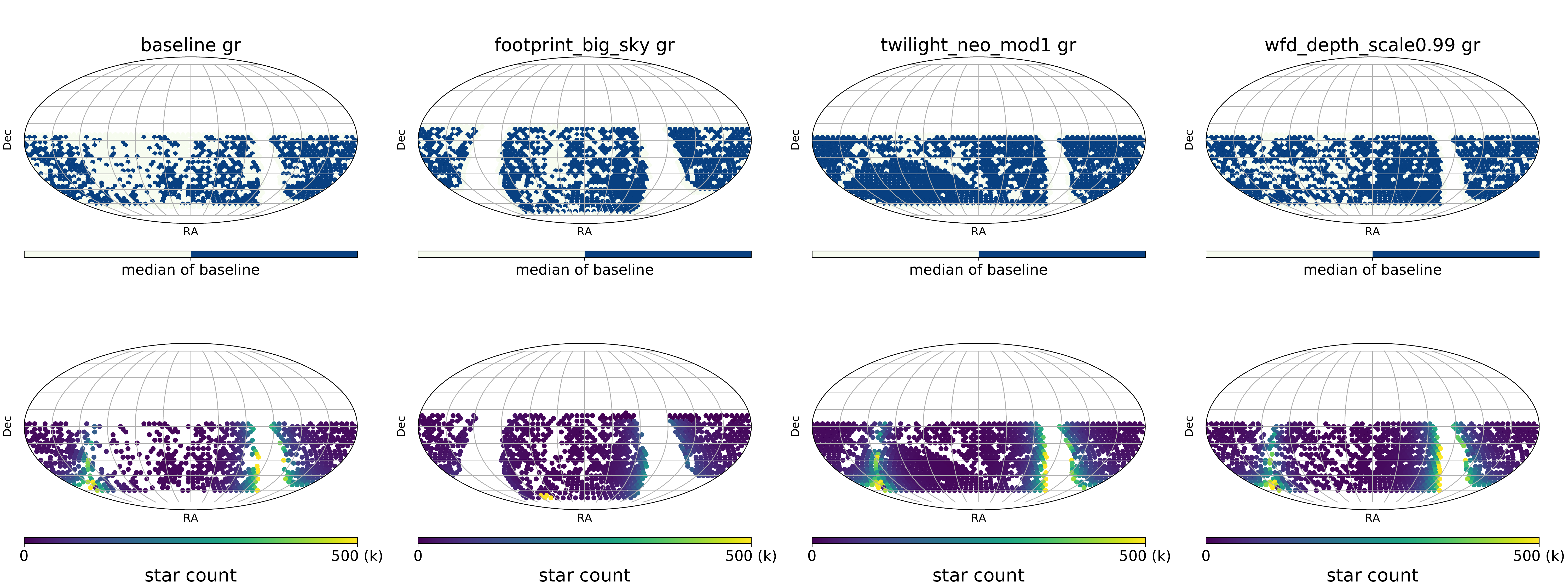}
\caption{This plot shows how the footprint figures of merit are calculated. The top four plots show the fields whose visit counts of $g-r$ pairs within a two-day interval are greater than the median counts in \texttt{baseline\_1.5}. The bottom four plots show the same but for the star count within those fields. The \texttt{footprint\_big\_sky} extends footprint to higher latitude than \texttt{baseline\_1.5}. The \texttt{twilight} allocates additional visits near twilight. See \autoref{sec:footprint}}
\label{fig:footprint_wfd}
\end{figure*}

\begin{figure*}
\centering
\gridline{
  \fig{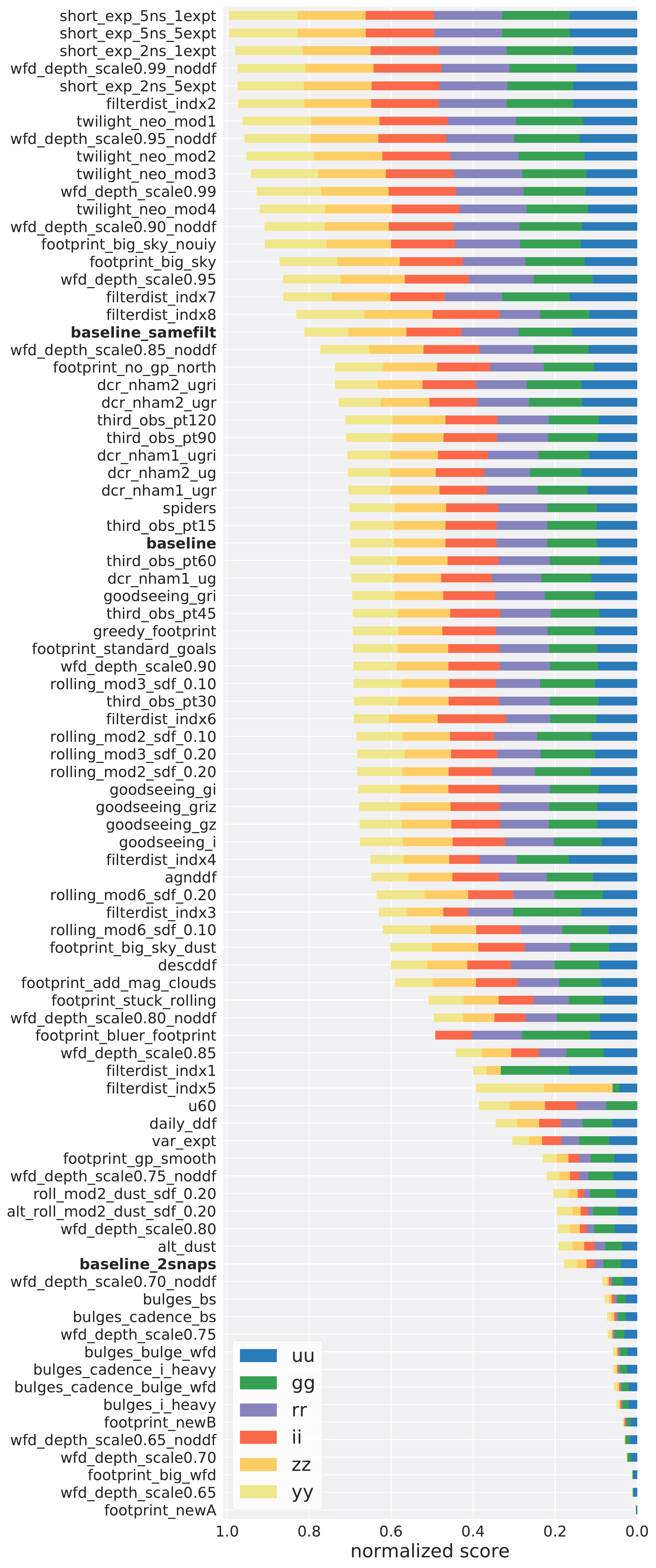}{0.4\textwidth}{($a$)} 
  \fig{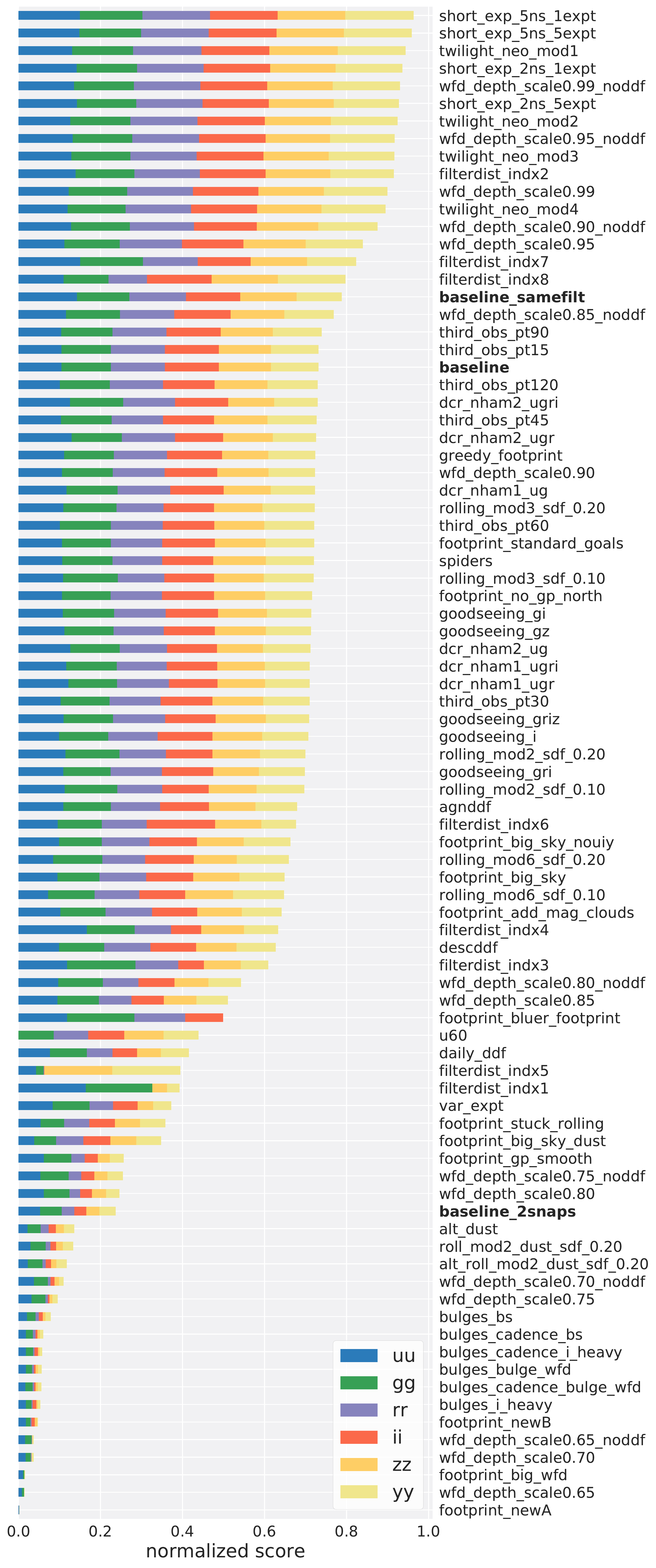}{0.4\textwidth}{($b$)}
 }
\caption{The figure of merit \fom$_\mathrm{EG}$ ($a$) and \fom$_\mathrm{Gal}$ ($b$) for all \opsim~ runs (for the WFD survey, selected as \texttt{proposalId = 1} in the \texttt{SQL} query, see \autoref{sec:method}) based on footprint coverage and star count with image pairs in the same filter (measuring lightcurve shape) as described in \autoref{sec:footprint} (\autoref{eq:fom:footprint}). {Colors and symbols denote filter-combinations using the same conventions as in Figure \ref{fig:barh_tgaps_wfd}} The two \fom s go hand in hand, with small differences in the ranking. 
}  
\label{fig:barh_footprint_same}
\end{figure*}

\begin{figure*}
\centering
\gridline{
  \fig{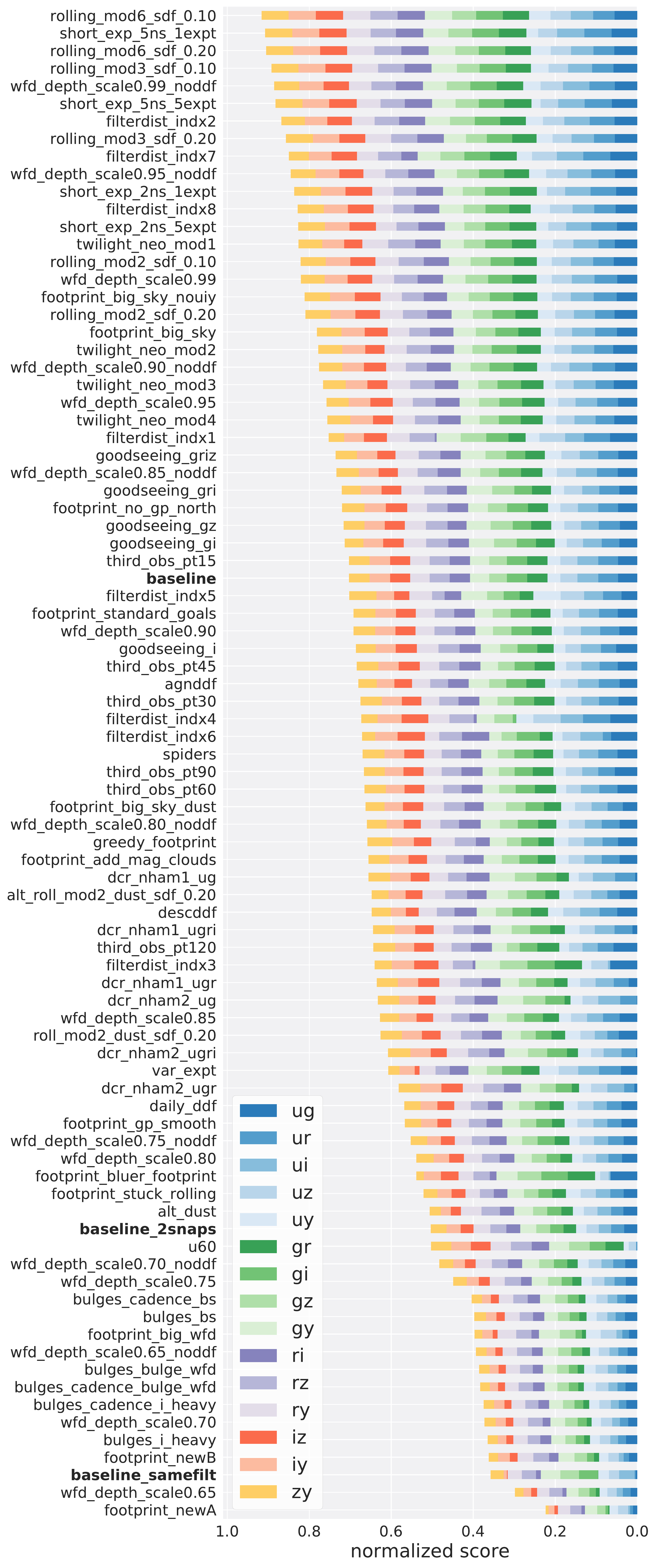}{0.4\textwidth}{($a$)}
  \fig{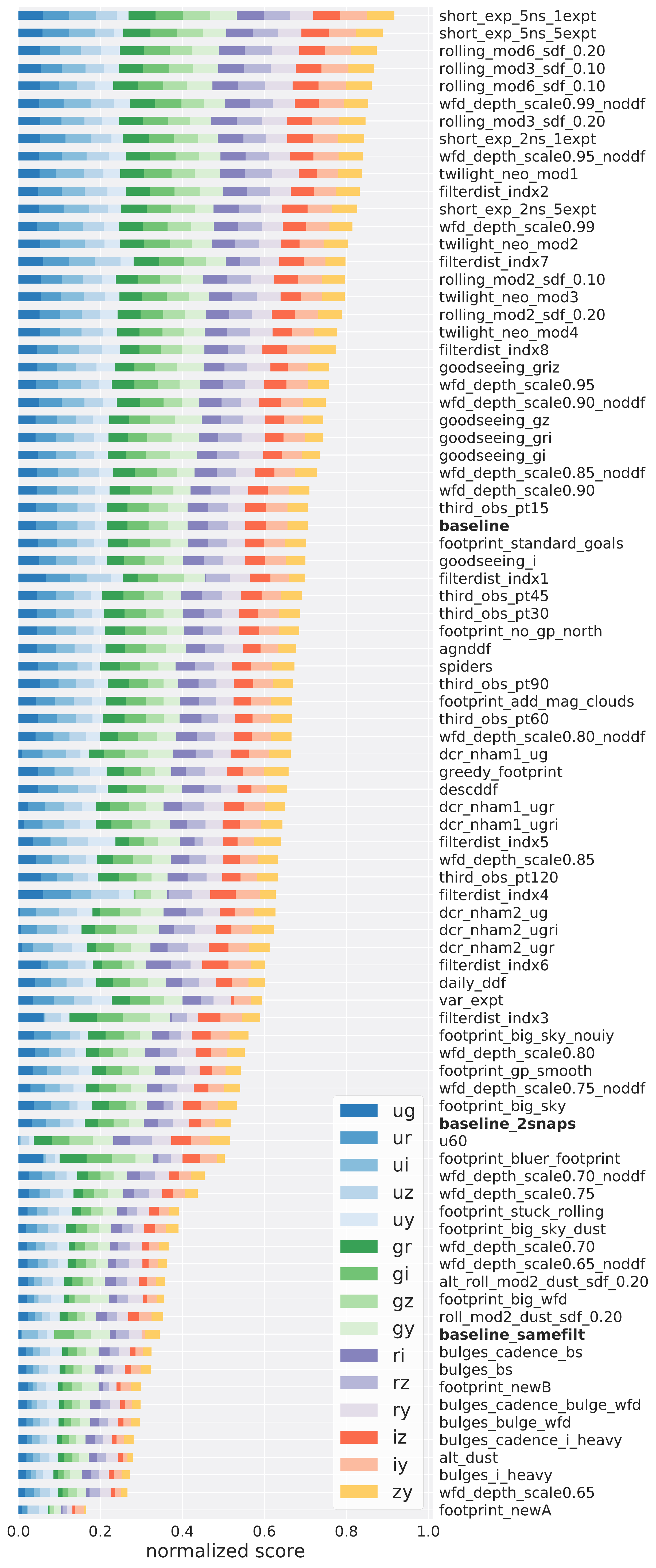}{0.4\textwidth}{ ($b$)} 
  }
\caption{As \autoref{fig:barh_footprint_same} but 
for image pairs in different filters (measuring color) as described in  \autoref{sec:footprint} (\autoref{eq:fom:footprint}). 
{Colors and symbols denote filter-combinations using the same conventions as in \autoref{fig:barh_tgaps_wfd}}. } 
\label{fig:barh_footprint_diff}
\end{figure*}

Footprint coverage is another important factor which plays a crucial role in determining LSST's ability to discover anomalous and unusual phenomena.

For the purpose of our analysis we define ``footprint'' as the extent of the sky (number of fields in the sky) that are ``well observed'' for each filter or filter-pair of interest. This approach is agnostic about the location of the fields in the sky, as we do not know where true-novelties may be. To decide if a field is ``well-observed,'' we compare the number of relevant observations 
to the number obtained in a chosen baseline LSST implementation (here, \texttt{baseline\_1.5}), under the motivation that the strategy ultimately adopted by the project should outperform this baseline.

In this context, ``relevant observations'' are defined slightly differently depending on whether one is measuring brightness evolution of color. For single filters that measure brightness evolution, all observations in that filter are relevant (so the comparison count is just the number of observations in the 10-year surveys). For filter pairs that measure the color, observations in a pair are only relevant if they occur within two days of each other (so the comparison count is the number of observation pairs constructed from images in different filters and collected within two days of each other). 

For the WFD in \texttt{baseline\_1.5} this results in thresholds for every filter pair listed in \autoref{table:WFD}. We acknowledge that this choice of threshold is somewhat arbitrary and that this will influence the result of this component of our \fom. We will return to the choice of threshold, and its impact on the science figures of merit, when we extend our analysis to other versions of the \opsim~strategies in \autoref{ss:v1.7}. {For the present, we emphasize that this thresholding is entirely relative to the baseline simulation: we are {\it not} imposing a requirement that the threshold guarantee a significant probability of detection.
Consider for example the $u-y$ filter pair: since there are not $u-y$ observations in the \texttt{baseline\_1.5} survey,\footnote{$u$ and $y$ filters benefit from very different sky conditions, and since the Rubin filter wheel can house only five out of the six filters at once, these two filters are likely to never be available in the same night.} a field with nonzero $u-y$~pairs would be considered ``well observed'' by us for that filter combination. 
By choosing a threshold relative to a fiducial implementation of the LSST survey, we seek to identify survey strategies that expand the potential of LSST.}


\begin{table}
\begin{tabular}{lllllll}
 &       u &     g &      r &      i &      z &      y \\
\hline
 u &  1711 &      &       &       &       &       \\
 g &    67 &  3570 &       &       &       &       \\
 i &    24 &    45 &    185 &  20582 &       &       \\
 r &    76 &   130 &  20301 &       &       &       \\
 z &     2 &    13 &     37 &    200 &  16470 &       \\
 y &     0 &     5 &     18 &     92 &    220 &  18431 \\
\hline

\end{tabular}
\caption{Thresholds for the footprint figures of merit based on \texttt{beseline\_v1.5} visit count for WFD observations. See \autoref{sec:footprint}.}
\label{table:WFD}
\end{table}

\begin{table}
\begin{tabular}{lllllll}
 &       u &     g &      r &      i &      z &      y \\
\hline
 u &  780 &      &      &      &      &      \\
 g &   58 &  1081 &      &      &      &      \\
 r &   40.5 &  60.5 &  1275 &      &      &      \\
 i &   11 &    19 &    46 &  1176 &      &      \\
 z &    1.5 &     8 &  11.5 &    67 &  1126 &      \\
 y &    0 &     4 &     8 &  30.5 &  60.5 &  1830 \\
\hline

\end{tabular}
\caption{Thresholds for the footprint figures of merit based on \texttt{beseline\_v1.5} visit count for Galactic Plane fields. See \autoref{ss:minisurveys}.}
\label{table:GP}
\end{table}
 
\begin{table}
\begin{tabular}{lllllll}
 &       u &     g &      r &      i &      z &      y \\
\hline
u &  684.5 &       &     &       &       &      \\
 g &   46 &  861.5 &     &       &       &      \\
 r &   37 &     47 &  882 &       &       &      \\
 i &    7 &   10.5 &   27 &  841.5 &       &      \\
 z &    1 &      2 &    6 &     67 &  924.5 &      \\
 y &    0 &      2 &    2 &   14.5 &   37.5 &  1458 \\

\hline

\end{tabular}
\caption{Thresholds for the footprint figures of merit based on \texttt{beseline\_v1.5} visit count for the LMC. See \autoref{ss:minisurveys}.}
\label{table:LCM}
\end{table}

\begin{table}
\begin{tabular}{lllllll}
 &       u &     g &      r &      i &      z &      y \\
\hline
u &  561 &     &     &     &     &      \\
 g &   32 &  741 &     &     &     &      \\
 r &   26 &   39 &  780 &     &     &      \\
 i &    5 &    9 &   18 &  861 &     &      \\
 z &    3 &    4 &    4 &   66 &  903 &      \\
 y &    2 &    4 &    5 &   19 &   40 &  1225 \\

\hline

\end{tabular}
\caption{Thresholds for the footprint figures of merit based on \texttt{beseline\_v1.5} visit count for the SMC. See \autoref{ss:minisurveys}.}
\label{table:SMC}
\end{table}
With these considerations in mind, the footprint figures of merit are generically calculated following the steps below. For each filter (or filter-pair) $k$:

\begin{itemize}
    \item count the number of visits for each filter pair; for same-filter pairs, consider all possible; for different-filter pairs, consider time gaps within 2 days;
    \item compute the median of this count in \texttt{baseline\_1.5} (call this $N_{\mathrm{median}, k}$);
    \item check if $N_k ~>~ N_{\mathrm{median}, k}$;
    \item sum over all fields that pass this requirement.
\end{itemize}

Depending on whether a scientist's focus is on extragalactic or  galactic anomalies, the preferred footprint would be different: for extragalactic anomalies one would simply want to maximize the sky coverage, whereas for Galactic science the probability of discovering an anomalous object or phenomenon would scale with the number of objects in the Galaxy in that observing field. Therefore, in addition to the \fom~just described, which focuses on extragalactic science and which we call  \fom$_\mathrm{EG}$ hereafter, we include one further footprint figure of merit, \fom$_\mathrm{Gal}$, that scales with the field's star density: \fom$_\mathrm{Gal}$ is the sum of each field that meets the requirements as described above, multiplied by the number of stars in that field (itself obtained from a realization of the \texttt{TRILEGAL} models of \citealt{girardi2005} accessed via \maf).

For an \opsim~ these \fom s are therefore defined as: 

\begin{eqnarray}
    p_{i,k} &=& 1 ~\mathrm{if} ~N_k ~>~ N_{\mathrm{median},k} ~\mathrm{else}~ 0 \nonumber\\ 
    FoM_\mathrm{EG} &=& \sum_k^{\mathrm{filters}} n_k  \sum_{i}^\mathrm{fields} p_{i,k}, \nonumber\\
    FoM_\mathrm{Gal} &=& \sum_k^\mathrm{filters} n_k \sum_{i}^\mathrm{fields} s_{i} p_{i,k}.
\label{eq:fom:footprint}
\end{eqnarray}
where $i$~is an index that ranges over all observed fields, $s_{i}$ is the star density (which is obtained from existing \maf~functions) for the $i$th field, and $p_{i,k}$ is set to 1 or 0 depending on whether the field meets the minimum visit requirement for that filter or filter-pair. 
Similarly to the depth figure of merit (\autoref{sec:depth}), the renormalization factor $n_k$~is the reciprocal of the maximum value (over all the \opsim s) of the sum over fields in the $k$'th filter. This renormalization serves to treat all the filters (or filter-pairs) on an equal footing: an \opsim~must be simultaneously top-ranked in all filters under consideration to achieve a \fom~value of 1.0. 
These figures of merit for all 86 simulations in \ovfive~ are plotted in  \autoref{fig:barh_footprint_same} and \autoref{fig:barh_footprint_diff}. 

While some \opsim s  were designed to cover a large footprint (such as \texttt{footprint\_bigsky}), other \opsim s perform better under the footprint figure of merit we develop here, which includes visit count thresholding in addition to simply evaluating the area covered.
So we see again the \texttt{short} and \texttt{rolling} cadences rising to the top.

\section{ Discussion}\label{sec:discussion}

We have created a series of \maf s and \fom s to assess the ability of Rubin Observatory LSST to discover completely novel astrophysical objects and phenomena. In an attempt to remain agnostic to what specific characteristic may render an object or phenomenon anomalous and thus which kind of anomalies we could discover, we choose to assess {the completeness of coverage achieved in a phase space quantified by figures of merit exploring the following observables}: 
{
\begin{enumerate}
    \item{Flux change, parameterized as \fom$_\mathrm{tGaps}$-magnitude;}
    \item{Color, parameterized as \fom$_\mathrm{tGaps}$-color;}
    \item{Depth, parameterized as \fom$_\mathrm{depth}$;}
    \item{Sky footprint, parameterized as \fom$_\mathrm{EG}$;}
    \item{Star counts, parameterized as \fom$_\mathrm{Gal}$.}
\end{enumerate}
}
(In contrast to \autoref{fig:barh_footprint_same} \& \autoref{fig:barh_footprint_diff}, in this Section the footprint and star-count figures of merit above are evaluated over all combinations of filter.)
The five elements enumerated above are added straightforwardly to one-another (\autoref{eq:fom:master}), although the final \fom~could be fine-tuned to some phenomenological expectations (for example to the discovery of \emph{galactic}{}, as opposed to \emph{extra-galactic} transients) by choosing the weights in the sum over the \fom~components. 

We note that the weights are thus formally somewhat arbitrary but scientifically rather crucial to the balance of science considerations imprinted on the sum figure of merit by the investigator.
Remaining  ``agnostic'', we opt to strive for balance in the normalization and relative weighting of each element of the \fom. The individual figures of merit are each normalized so that they essentially rank all the \opsim s on a 0-1 scale for that particular dimension in feature space, where an \opsim~must be top-ranked simultaneously in each filter (or filter-pair) to achieve a maximum 1.0 score (Sections \ref{sec:timegaps} - \ref{sec:footprint}). We then choose weighting factors ($w_i$~in \autoref{eq:fom:master}) to weight each of the five \fom s equally.



\subsection{Main Survey} \label{ss:mainsurvey}

\begin{figure*}
\centering
\gridline{
  \fig{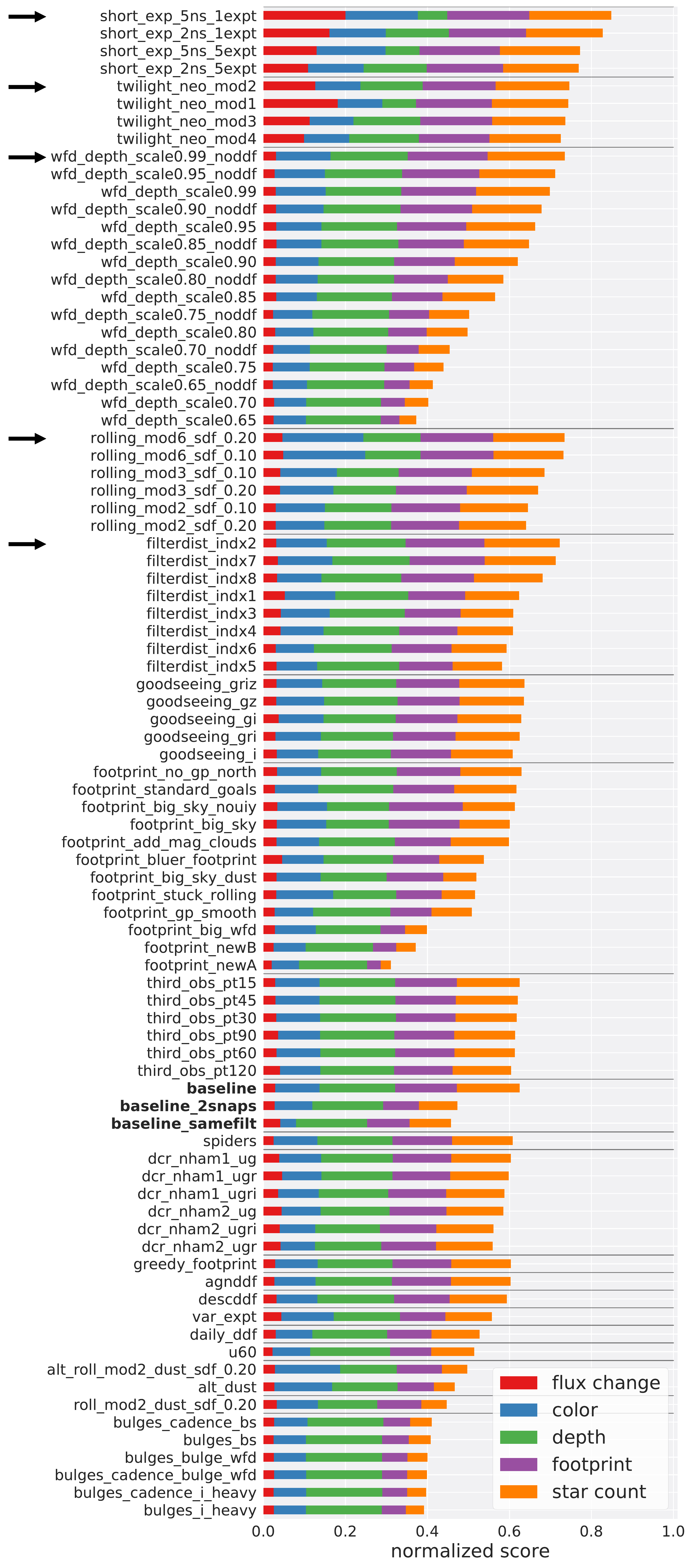}{0.4\textwidth}{($a$)}
  \fig{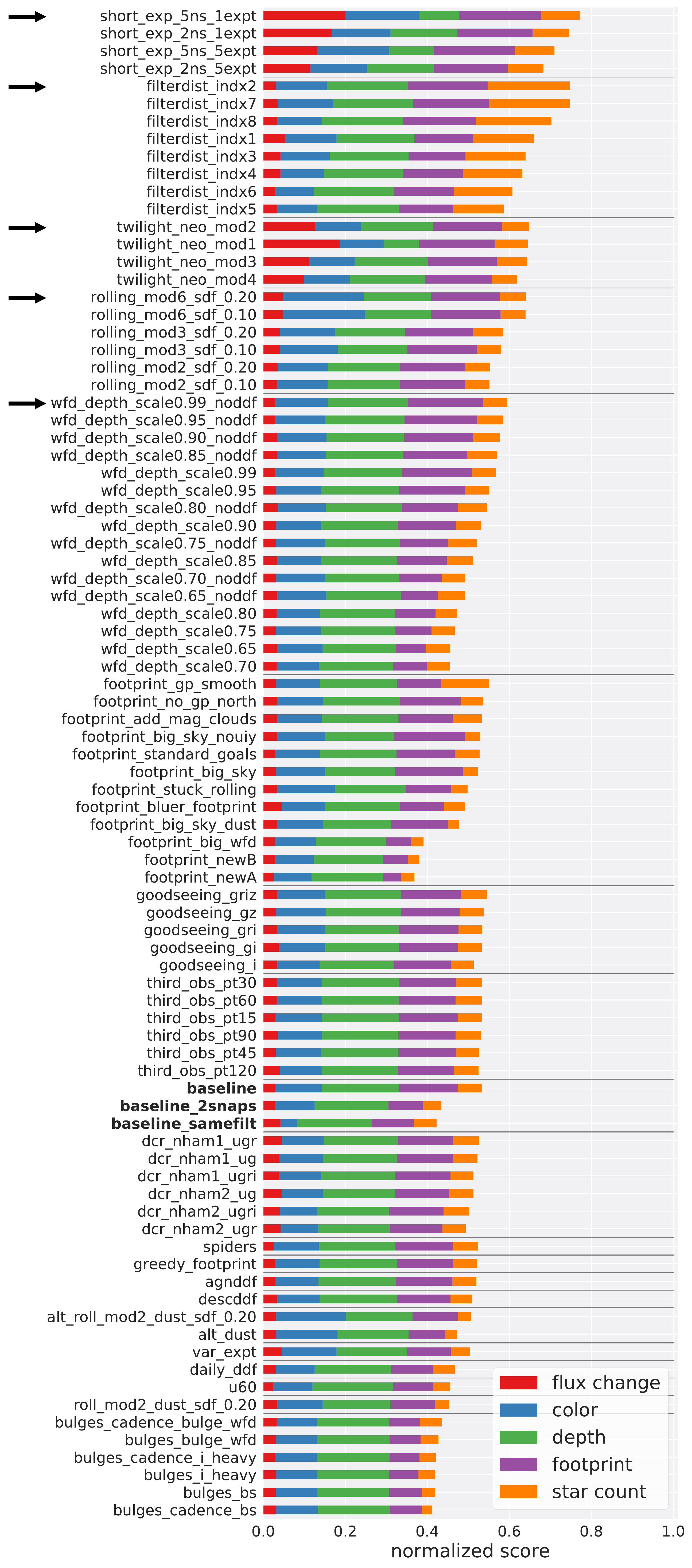}{0.4\textwidth}{($b$)}
}
\caption{{Bar plot showing the performance for our final five-fold \fom} ranked by family's top performing \opsim: ($a$) for WFD observations selected by setting \texttt{proposalId=1}; ($b$) all observations not identified with DDF. Arrows point to the \opsim s that are also shown in the radar plots in \autoref{fig:radar_wfd}. This plot is discussed in
\autoref{sec:discussion}. 
}
\label{fig:barh_wfd}

\end{figure*}

\begin{figure*}
\centering
\gridline{
  \fig{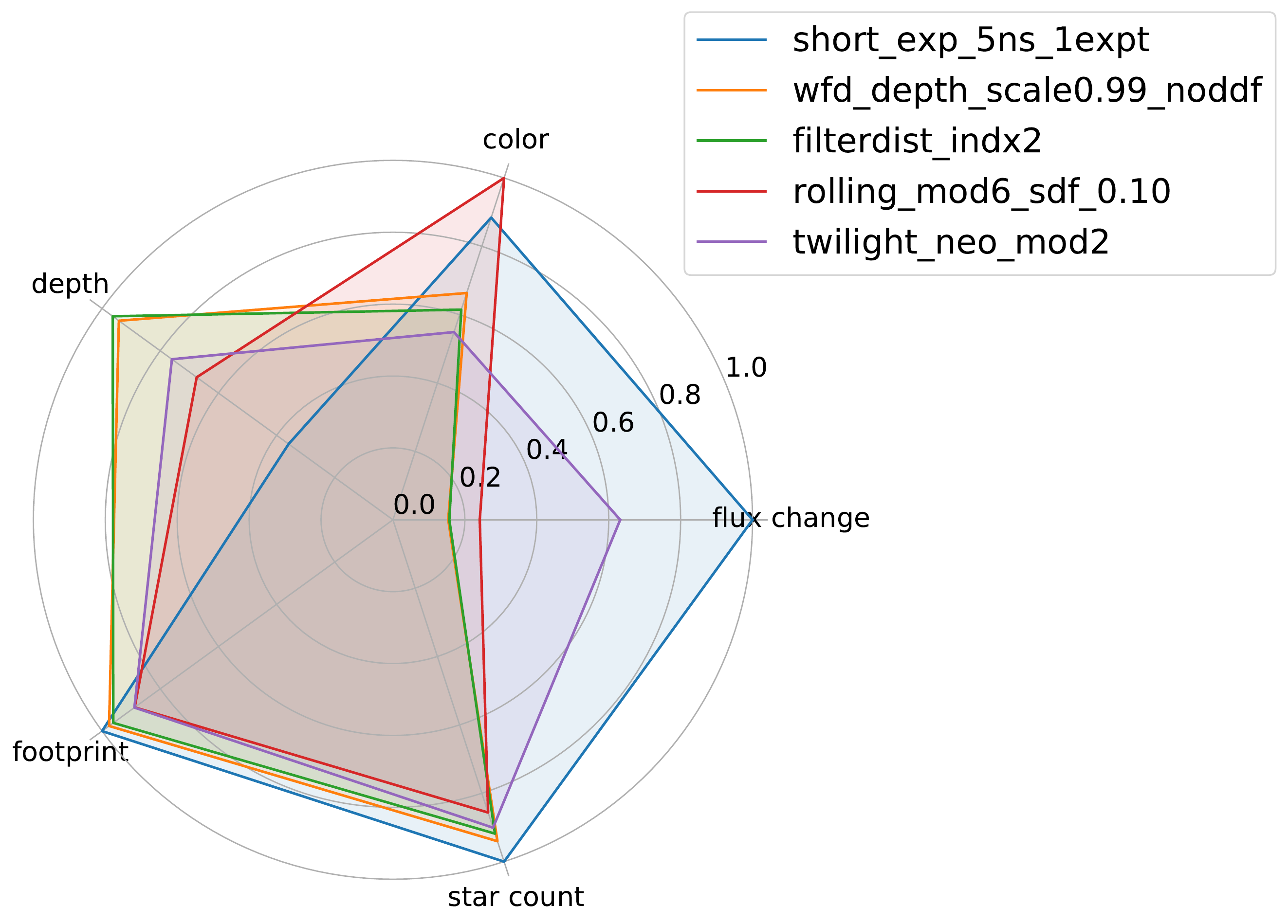}{0.5\textwidth}{($a$)}
  \fig{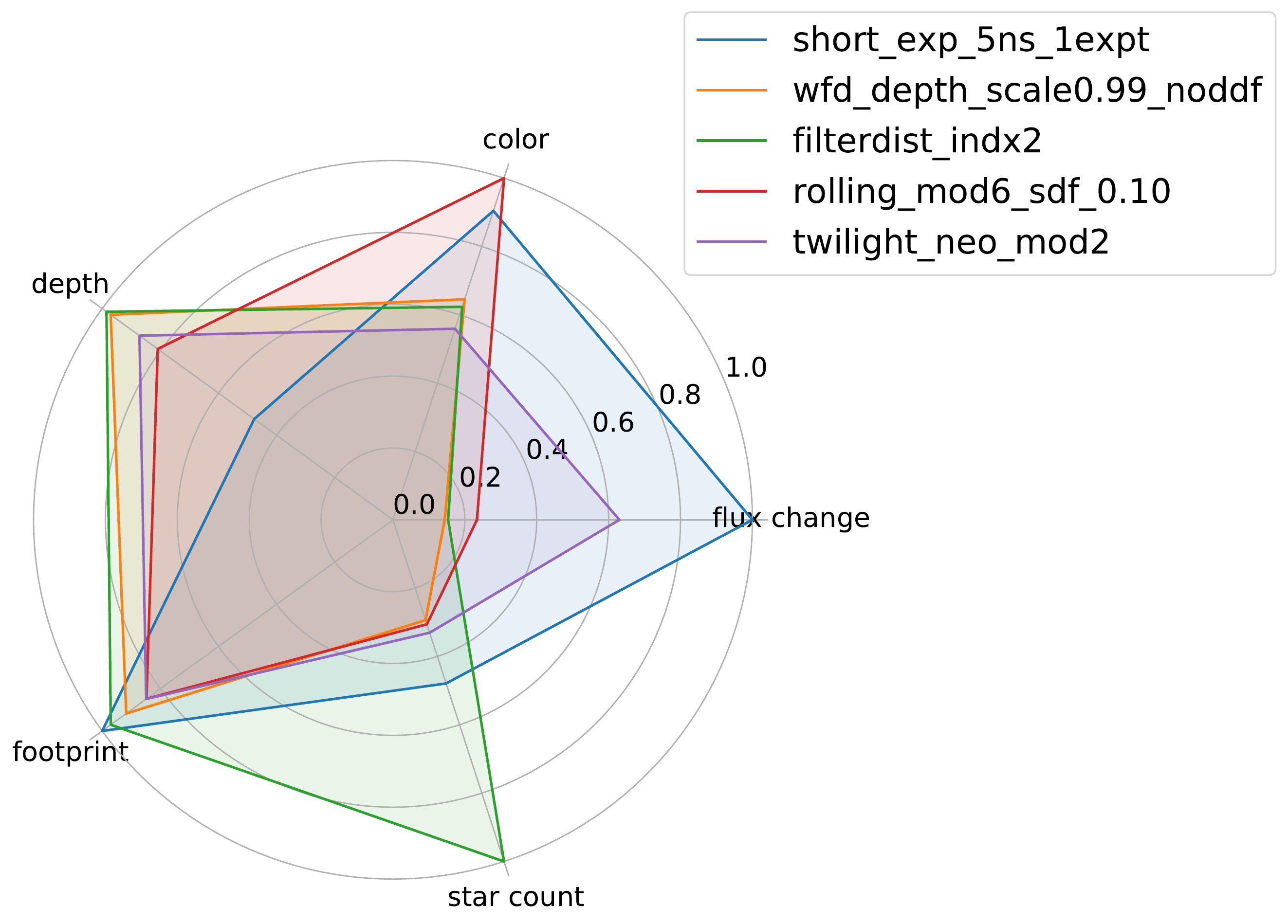}{0.5\textwidth}{($b$)}
}
\caption{The radar plot the highest performing \opsim~in each of the top five \opsim~families: ($a$) for the WFD survey (\texttt{proposalId=1}); ($b$) all region excluding DDFs. 
An interactive version of this plot is available at \url{https://xiaolng.github.io/widgets/radar.html}, see \autoref{appendix}. A radar plot shows the metrics at the vertices of an polygon with the metric value mapped to the distance from the center of the polygon. With multiple \opsim s plotted in the same radar plot, we can compare the tensions between \fom ~components, while the total area inside of the polygon is a measure of the overall quality of the \opsim. For our \fom, which is the simple sum of five components, this visualization is well suited to provide a synoptic view.
 This plot is discussed in \autoref{ss:mainsurvey}}\label{fig:radar_wfd}
\end{figure*}

First, we want to summarize some considerations arising from our analysis of the performance of different LSST simulations from  \ovfive~. These considerations, however, should be read in the light of the discussion of the different \opsim~versions in \citet{frontpaper} and rely on the reader having familiarity with the \ovfive~set, as described there in and in more detail on the Rubin Community forum\footnote{\url{https://community.lsst.org/t/fbs-1-5-release-may- update-bonus-fbs-1-5-release/4139}}.
We will also extend this discussion to other versions of \opsim s briefly in \autoref{ss:v1.7}.

The bar charts included in this work (\autoref{fig:barh_tgaps_wfd}, \autoref{fig:barh_depth}, \autoref{fig:barh_footprint_same}, \autoref{fig:barh_footprint_diff}) provide an intuitive way to understand how sensitive a \fom ~is to observing cadence choices. We note that: 
\begin{itemize}
\item The \fom$_\mathrm{tGaps}$ for flux evolution (\autoref{fig:barh_tgaps_wfd} $a$) is very sensitive to \opsim~ details: \opsim s that included short exposures are critically improved as they provide visibility into time-scales that are otherwise not accessible to the survey. Going back to the the  phase space of transients presented in  \autoref{fig:phasespace} and the discussion of existing observational biases, rapid evolutionary time scales are quite likely to host unobserved, unexpected phenomena: \emph{true-novelties}. Our metric reflects this expectation. 
\item This effect is mitigated by the depth metric that down-weights \opsim s where the short exposures come at a cost of overall survey depth. Otherwise, this metric does not differ much across most \opsim s as the median observation depth is well defined by the \citet{lsst}. 
\item The galactic and extra-galactic footprint metrics as defined by us are somewhat less sensitive to observing choices, as indicated by the more gentle slope of the silhouette of the bar chart in \autoref{fig:barh_footprint_same} and \autoref{fig:barh_footprint_diff}. However, in \autoref{fig:barh_footprint_same}  three regimes are visible: \opsim s that include short exposure (\texttt{twilight}, \texttt{short\_exp}, and some \texttt{wfd\_depth} implementations with a large fraction of observations included in the WFD survey, 
\ie~a large ``scale'' parameter) raise to the top. A number of specific implementations from nearly all families, however, sink to the bottom and perform very poorly (some footprint implementation and \texttt{wfd\_depth} surveys with small scale parameter).
The ranking of the \opsim s is similar for \autoref{fig:barh_footprint_same} and \autoref{fig:barh_footprint_diff}.

\end{itemize}

\autoref{fig:barh_wfd} shows the performance for the combined \fom~ as described above, organized by \opsim~ family. Observations associated with the WFD proposal are shown in   \autoref{fig:barh_wfd}-$a$ (the results including the minisurveys are shown in panel $b$; the mini-surveys themselves are discussed in more detail in  \autoref{ss:minisurveys}). This visualization provides a synoptic look at our \fom. Individual components of the \fom ~can still be identified by the color of the bar element. Furthermore, this visualization allows us to identify the performance of a family of \opsim s, providing a more intuitive way to assess the reason why \opsim s may rank differently, but also a way to assess how the detail of an implementation affect results. For example, the \texttt{short} and \texttt{twilight} families are among the top performers, with little sensitivity to the details of the implementation. Conversely, the \texttt{wfd} and \texttt{footprint} families (the former ranking third overall, the latter in the middle ranking seventh)  provide a range of results, from excellent to poor, depending on the implementations details. In both cases, the performance is dominated by the footprint \fom~ (the purple portion of the bar. For the \texttt{wfd} the result of both footprint \fom s scales with the ``scale'' parameter, the number of visits allocated to the WFD survey, but see also \autoref{ss:minisurveys}). It should be noted that these are core families of simulations, with a range of implementation details that can be tweaked, so it is not surprising that they result in a range of measured performance. See \autoref{ss:opsim}, \autoref{tab:opsim}, and  \citet{frontpaper}   for more detail.

In \autoref{ss:minisurveys} we will discuss \autoref{fig:barh_wfd}-$b$ and address the question of what the minisurveys add to the science performed in the WFD regions, by considering together all the exposures not identified with a deep-drilling field.

Applying our metrics to \ovfive, we note that: 
\begin{itemize}
\item
The \fom$_\mathrm{tGaps}$-magnitude-evolution component (see also \autoref{sec:timegaps}) is pushing entire families of \opsim s to the top, namely those that include short observations and thus expand the LSST feature space to short time scales. 
\item
Within an \opsim~ family, the most significant contribution in determining the ranking of \opsim s are the \fom$_\mathrm{EG}$ and \fom$_\mathrm{Gal}$ that are, however, strongly correlated (see also \autoref{sec:footprint}).  

\item Overall, the top performing \opsim s in each family are all within a score of $\sim0.3$ of each other, demonstrating that all \opsim~ families have the potential of being implemented in a way that is favorable to the discovery of true-novelties, with the exception of specialized surveys such as \texttt{bulge}, and \texttt{alt\_dust}: these families that typically allocate visits to focus areas of the sky are penalized in the footprint portion of our \fom. We refrain from discussing the \texttt{rolling} family of \opsim s until \autoref{ss:v1.7}.
\end{itemize}
A radar plot in \autoref{fig:radar_wfd} shows selected \opsim s to tune the balance in the design of the final strategy. In \autoref{fig:radar_wfd} the best performing \opsim s for each of the top four families are plotted as identified above (panel $a$ for the WFD). With this visualization, we can see the substantial impact of the flux-change component of the metric, which measure completeness in pairs of observations in the same filter, on the overall result, and how it is compensated, in the case of \texttt{short\_exp} by the depth \fom$_\mathrm{depth}$, leaving the \texttt{twilight\_neo\_mod2} as the most balanced \opsim for our set of metrics. 

We provide an interactive widget that allows the reader to explore  the radar plot for our and other sets of metrics 
in \autoref{appendix}.


\subsection{Minisurveys}\label{ss:minisurveys}
In addition to the primary WFD survey, LSST has the capability of conducting mini-surveys including but not limited to the Galactic Plane, Magellanic Clouds and Deep Drilling fields. These minisurveys enhance science cases that yield greater science return with greater density of targets, including (but not limited to):
the detection of stellar-mass black holes, dwarf novae and Type Ia Supernova progenitors, and gravitational microlensing at various timescales. 
Because the mini-survey regions tend to cover regions of high density of stellar sources, they are therefore more likely to discover phenomena never observed before. 

To assess the coverage achieved in areas of interest to the minisurveys, we select observations by spatial footprint (rather than by \texttt{proposalID}), as discussed in \autoref{sec:method}. 

\begin{figure*}[t!]
\centering
\includegraphics[scale=0.5]{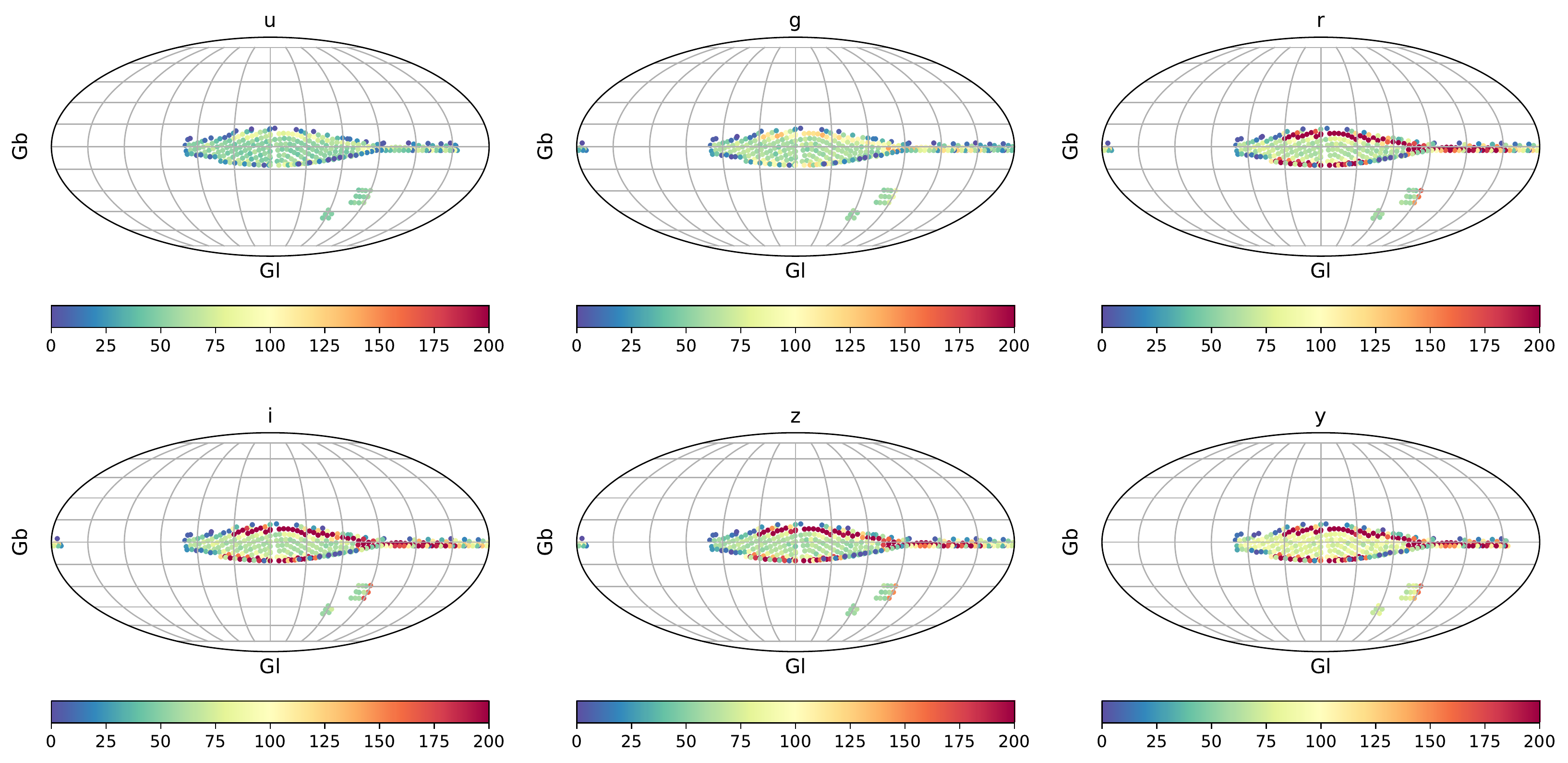}
\caption{ The footprint of the Galactic Plane and Magellanic Clouds. The definition of these spatial regions and the selection of the corresponding footprint are described in \autoref{ss:minisurveys}. We select 254 fields for the Galactic Plane, 12 fields for the LMC and 5 fields for the SMC. The color shows the number of visits in \texttt{baseline\_v1.5} in six bands in each field. 
}
\label{fig:footprint_ms}
\end{figure*}

\autoref{fig:footprint_ms} shows the adopted minisurvey regions: Galactic Plane (GP), Large Magellanic Cloud (LMC), and Small Magellanic Cloud (SMC). The adopted GP footprint is a cosine function of Galactic longitude, with amplitude $|b| = 10^\circ$~and first zero at $|l| = \pm 85^\circ$, {plus a strip at constant thickness $b \leq 2.5^{\circ} $~to accommodate the thin disk at all longitudes.}\footnote{{This Galactic Plane footprint is similar to the ``zone of avoidance'' from high-density Galactic Plane regions defined in method {\tt \_plot\_mwZone()} in the \maf~module {\tt spatialPlotters.py}, except there the first zero occurs at $l = \pm 80^{\circ}$.}}
For the Magellanic Clouds, we select all healpix fields (resolution parameter \texttt{NSIDE}=16) within 3.5 FOV of the twelve fields covering the cloud main bodies proposed in the \citet{olsenMCs} cadence whitepaper. 

We also provide code for the user to choose a specific region of the sky of their interest (see \autoref{appendix}) either by setting a formula from coordinate parameters, or interactively selecting pixels. 

{The individual figures of merit in these regions are normalized following similar schemes as for the main survey, but with thresholds or maximum values evaluated over the spatial regions of interest (see Tables \ref{table:GP}-\ref{table:SMC} for the comparison $N_{\mathrm{median},k}$~counts for the minisurvey spatial regions).} 
We normalize the \fom ~of the time gap metric by its maximum value across the \opsim s. For the footprint, as with \autoref{sec:footprint}, we choose the median number of visits from \texttt{baseline v1.5} within the defined footprint, normalized by the total selected number of fields within (254 for Galactic Plane, 12 for the LMC and 5 for the SMC), as a threshold to decide whether to classify a field as ``well observed'' (Tables \ref{table:GP} - \ref{table:SMC}). 

\begin{figure*}
\centering
\gridline{
  \fig{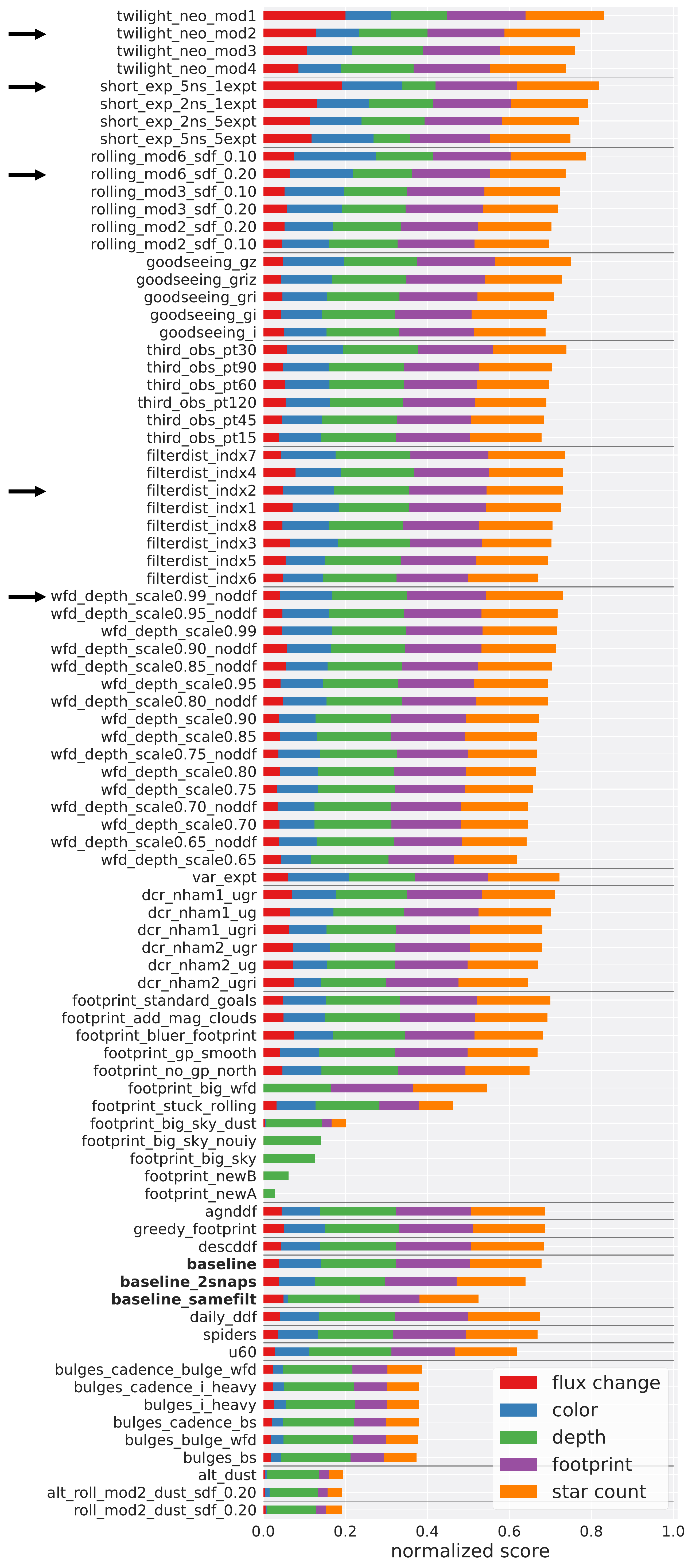}{0.3\textwidth}{($a$)}
  \fig{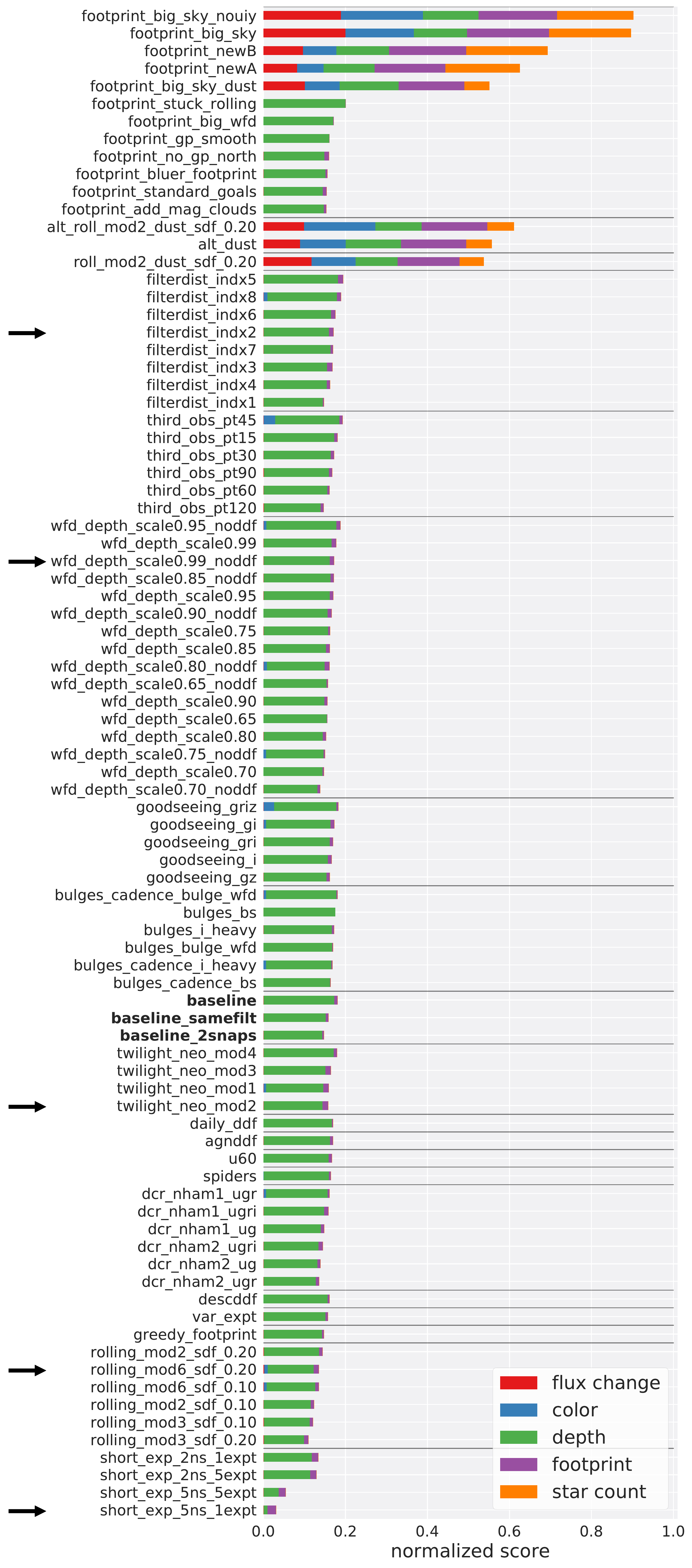}{0.3\textwidth}{($b$)}
\fig{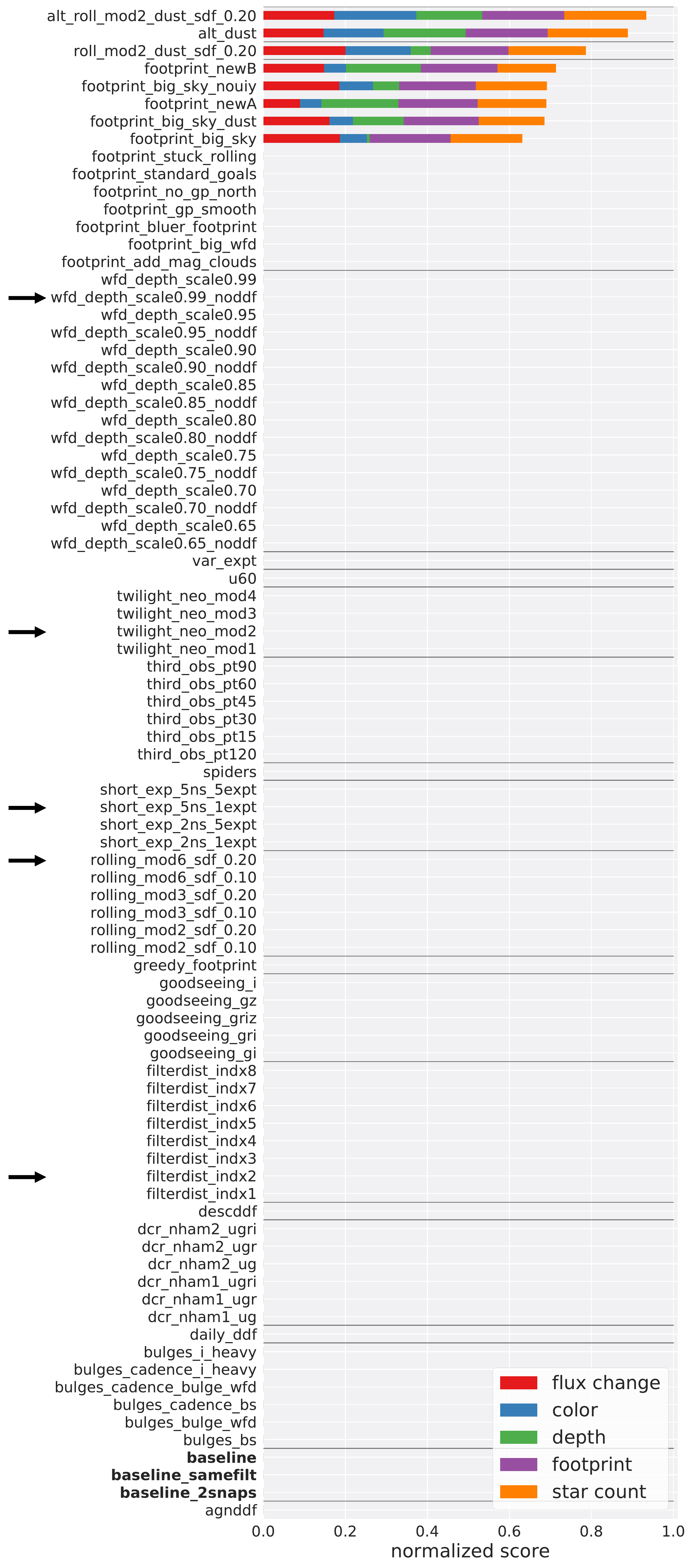}{0.3\textwidth}{($c$)}
}
\caption{Bar plot, as \autoref{fig:barh_wfd}, showing \opsim  s ranked by family, but this time for three selected spatial regions:  ($a$) the Galactic Plane ($b$) the LMC ($c$) the SMC. {Only visits allocated to the WFD (labeled as \texttt{proposalId=1}) are counted}. Arrows point to the \opsim s that are also shown in the radar plots in \autoref{fig:radar_wfd}.  The \texttt{twilight} and \texttt{short} families of \opsim perform best on the Galactic Plane, as they did over the entire WFD footprint, while \texttt{wfd\_depth}, formerly ranked third, is now ranked the seventh. But in reality the top performing \opsim~ in most families all perform similarly. The main differences are generally driven by  \fom$_\mathrm{tGaps}$ in the same filter. Only eight \opsim s cover the SMC and only five cover both LMC and SMC with WFD-identified observations.  See \autoref{ss:minisurveys}. }
\label{fig:barh_ms_wfd}
\end{figure*}

\begin{figure*}
\centering
\gridline{
  \fig{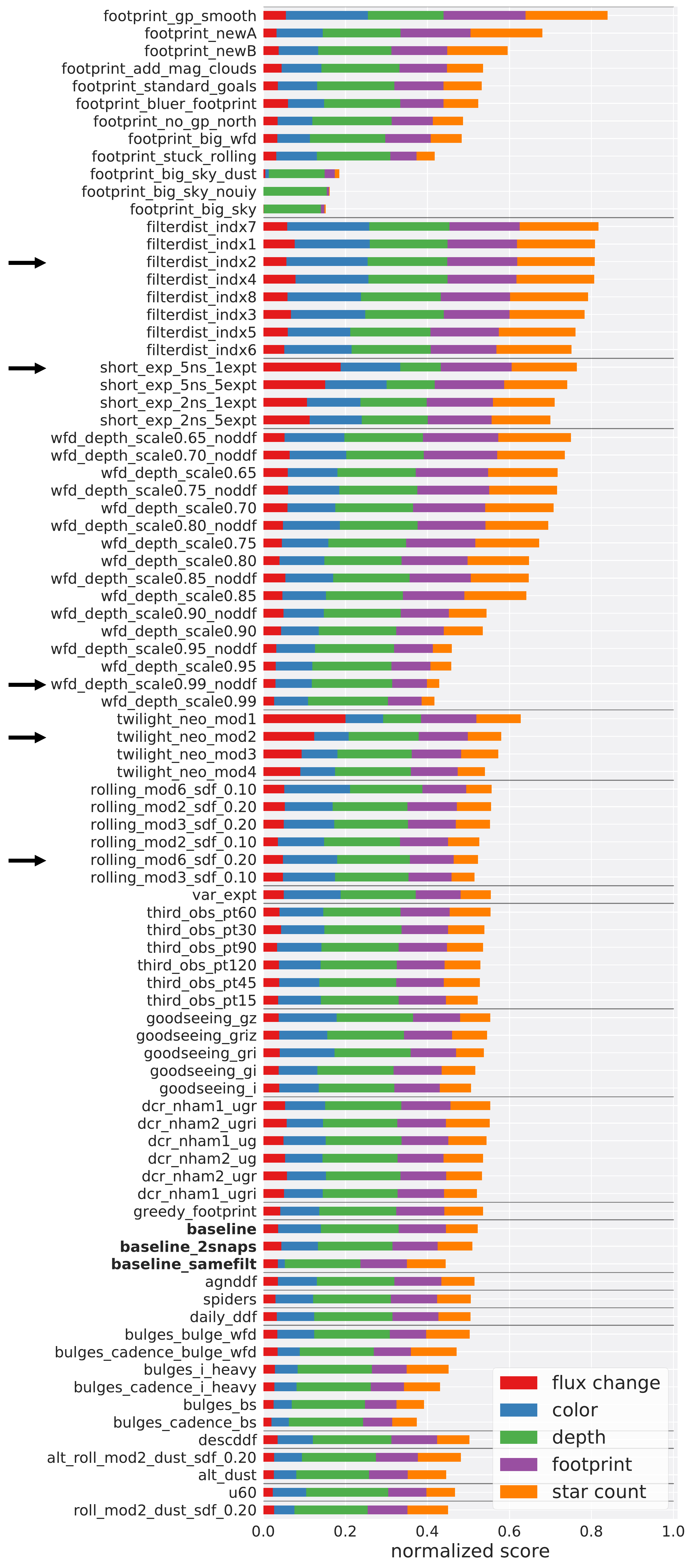}{0.3\textwidth}{($a$)}
  \fig{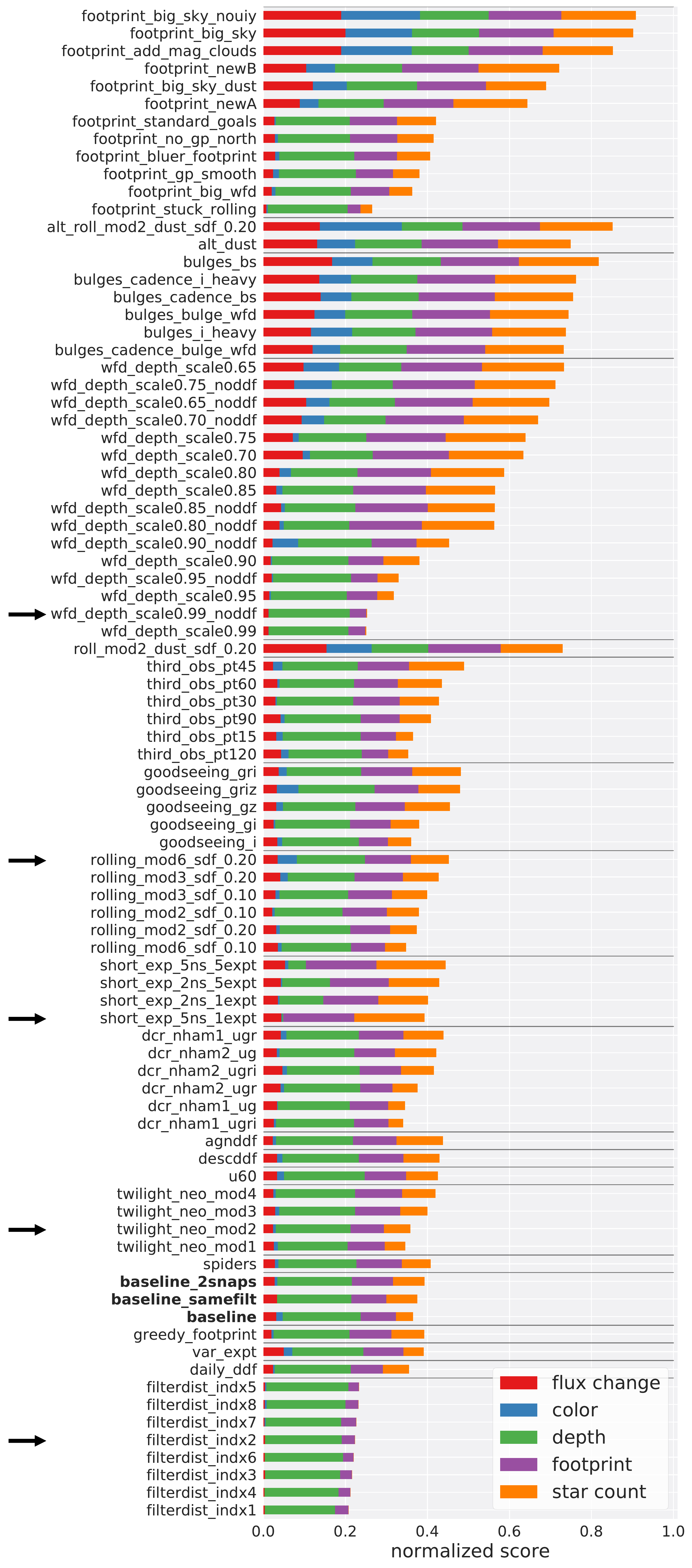}{0.3\textwidth}{ ($b$)}
\fig{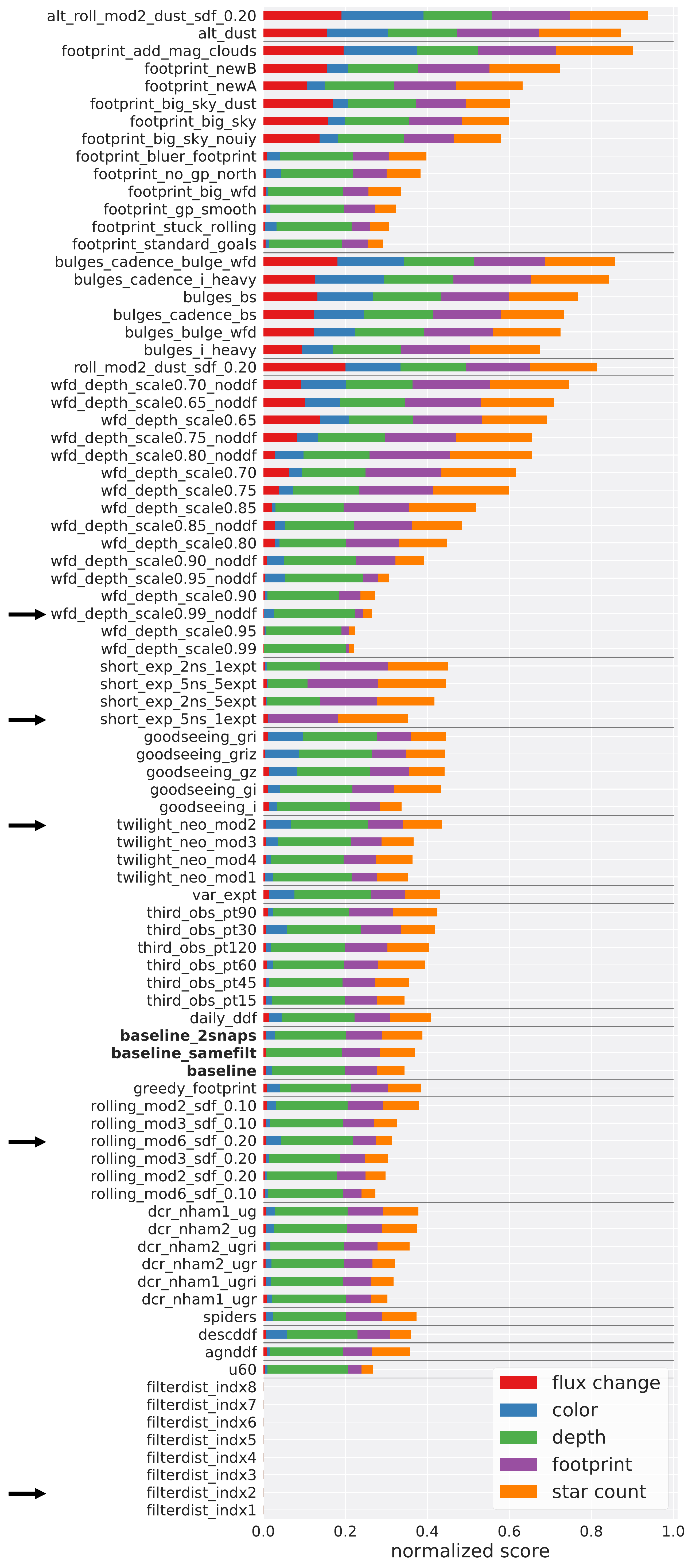}{0.3\textwidth}{($c$)}
}
\caption{As \autoref{fig:barh_ms_wfd} but with all visits excluding DDFs being counted for the minisurvey spatial regions: ($a$) Galactic Plane ($b$) LMC ($c$) SMC. Arrows point to the \opsim s that are also shown in the radar plots in \autoref{fig:radar_wfd}. While the \texttt{footprint} family has the best performing \opsim~overall for the Galactic Plane (\texttt{footprint\_gp\_smooth}) it also has the least performing \opsim s, showing the largest dynamical range due to the combined effects of both the footprint and time-gap metrics. Meanwhile, all \texttt{filterdist} \opsim s perform well. 
However, the LMC and SMC are now covered with most \opsim s with the exception of the \texttt{filtdist} family. See \autoref{ss:minisurveys}.}
\label{fig:barh_ms_all}
\end{figure*}

{\autoref{fig:barh_ms_wfd} \& \autoref{fig:barh_ms_all} present the evaluations of the figures of merit on the minisurvey regions. \autoref{fig:barh_ms_wfd} shows the figure of merit evaluation for the three spatial regions for fields that are allocated WFD coverage (\ie,  observations with \texttt{proposalID=1}). This demonstrates quite dramatically that the Magellanic Clouds are not allocated WFD-like coverage in most of the strategies considered.}
{ \autoref{fig:barh_ms_all} widens the evaluation to include any exposures not associated with deep drilling fields. Strong variation is apparent between the families of \opsim s, as expected for families that experiment with the areas of coverage on-sky. The \texttt{footprint} family shows strong variation depending on which region is favored: \texttt{footprint\_gp\_smooth} performs the best for the Galactic Plane, but is in the bottom quartile (of all the \opsim s) for the Magellanic Clouds. Conversely, \texttt{footprint\_add\_mag\_clouds} is at or near the top for the Magellanic Clouds, but is near the middle for the Galactic Plane regions. 

The \texttt{AltSched} implementations perform quite badly for the Galactic Plane regions, but allocate favorable observations to the Magellanic Clouds. Curiously, the \texttt{bulges} family of \opsim s are among the {\it worst}-performing families for the Galactic Plane regions, though in the top three for the Magellanic Clouds. 
The baseline strategies appear near the middle of the distribution for the minisurvey regions.} 

{Since the regions are to some extent competing with each other in terms of allocation, \autoref{fig:barh_ms_wfd} \& \autoref{fig:barh_ms_all} may be best interpreted in terms of which \opsim s to avoid due to their being problematic for particular regions of scientific importance. From that perspective, \opsim s \texttt{alt\_dust} and \texttt{footprint\_new} are unlikely to satisfy those interested in the Galactic Plane, while the \texttt{filterdist} family serves the Magellanic Clouds particularly poorly. } 

By comparing \autoref{fig:barh_ms_all} panel $a$ and $b$ we can  address the question of what the minisurveys add to the science performed in the WFD regions, by considering together all the exposures not identified with a deep-drilling field (\autoref{fig:barh_ms_all}-$b$). Most of the \fom s remain relatively unchanged by the inclusion of the minisurvey exposures. The exception is the ``star density''~\fom, which appears to return systematically {\it lower} values for each \opsim~when the minisurveys are {\it included}. This is probably an artefact of normalization and thresholding: \opsim~s that well-cover regions of high stellar density will increase the upper bound of the range of this \fom~across the \opsim s. If one or two \opsim s stand out from the rest in this regard, the standouts will return renormalized \fom~near 1.0 while the rest sink to lower values.
{Inclusion of the minisurvey-identified observations in the \fom~computation changes the ordering of the families somewhat, though not radically: the baseline strategies, for example, remain in the bottom quartile of the \opsim s when ranked by family (see panel $b$ of \autoref{fig:barh_ms_all}). The minisurveys do seem to reduce the contrast somewhat between \opsim s within a given family and even between the families.}



\subsection{Comparison with v1.7}\label{ss:v1.7}
Our work is based on \ovfive, the version of \opsim~ simulations released in May 2020. However, since then, more simulations have been released. We briefly inspected the performance 
of \ovseven~ (74 simulations at the time of writing) and \ovsevenone~ (10 simulations), the most recent simulations at the time of writing. 

It is important to note some key differences between \ovfive,  \ovseven, and \ovsevenone~ (however, a thorough description of these simulations is outside the scope of this paper and the reader is reminded that details are available on the Rubin Community web forum.\footnote{
\ovfive~ \url{https://community.lsst.org/t/fbs-1-5-release-may-update-bonus-fbs-1-5-release/4139}, \\
\ovseven~ \url{https://community.lsst.org/t/survey-simulations-v1-7-release-january-2021/4660},\\
\ovsevenone~ \url{https://community.lsst.org/t/survey-simulations-v1-7-1-release-april-2021/4865}})

\ovfive~ uses for almost all simulations 1$\times$30 seconds exposures while \ovseven~and \ovsevenone~ use 2$\times$15 seconds exposures per visit. It estimated that this would lead to a loss of efficiency of $\sim7\%.$\footnote{see for example \url{https://community.lsst.org/t/october-2019-update-fbs-1-3-runs/3885}} It is also expected that the \texttt{rolling} family of \opsim s would display significant changes compared to \ovfive, due to improvements in the way rolling cadences are implemented to more closely match their specifications. Versions \texttt{1.7} and later of the \texttt{rolling} \opsim s are considered a more reliable implementation of rolling cadence than \texttt{v1.5} (Lynne Jones, private communication).

\begin{figure*}
\centering
\gridline{
  \fig{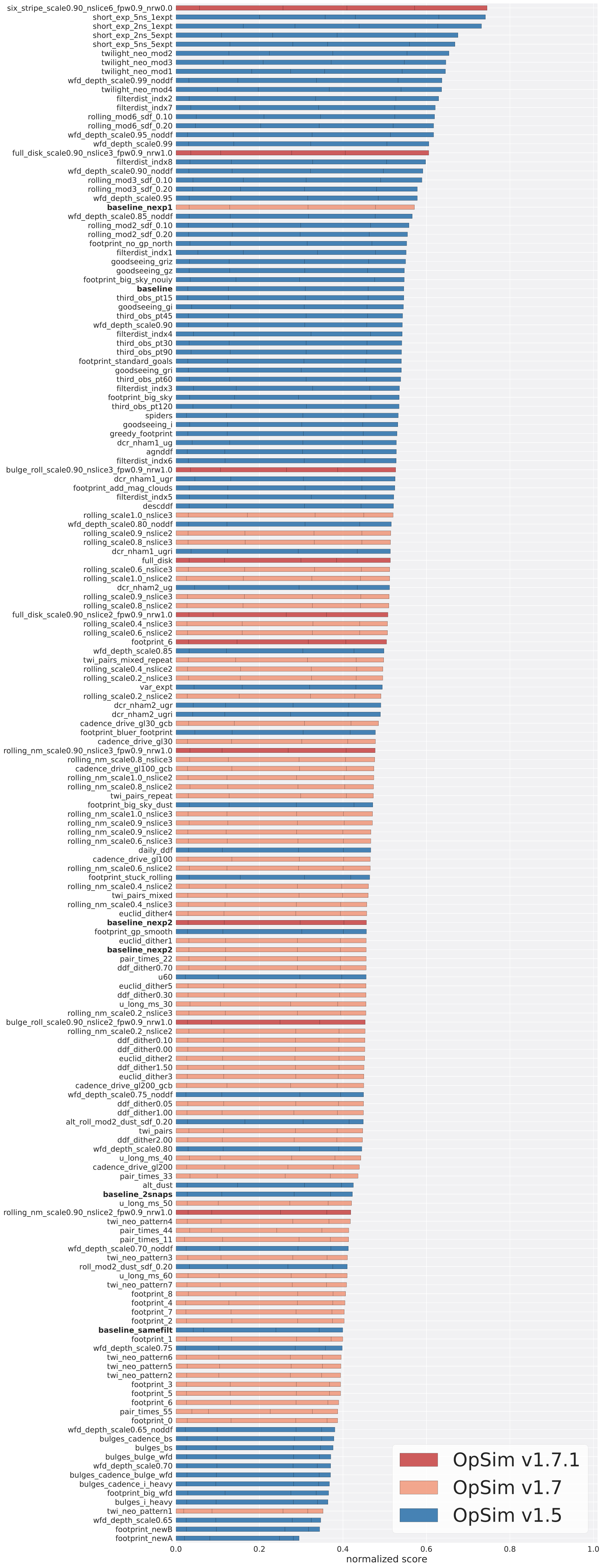}{0.4\textwidth}{($a$)}
  \fig{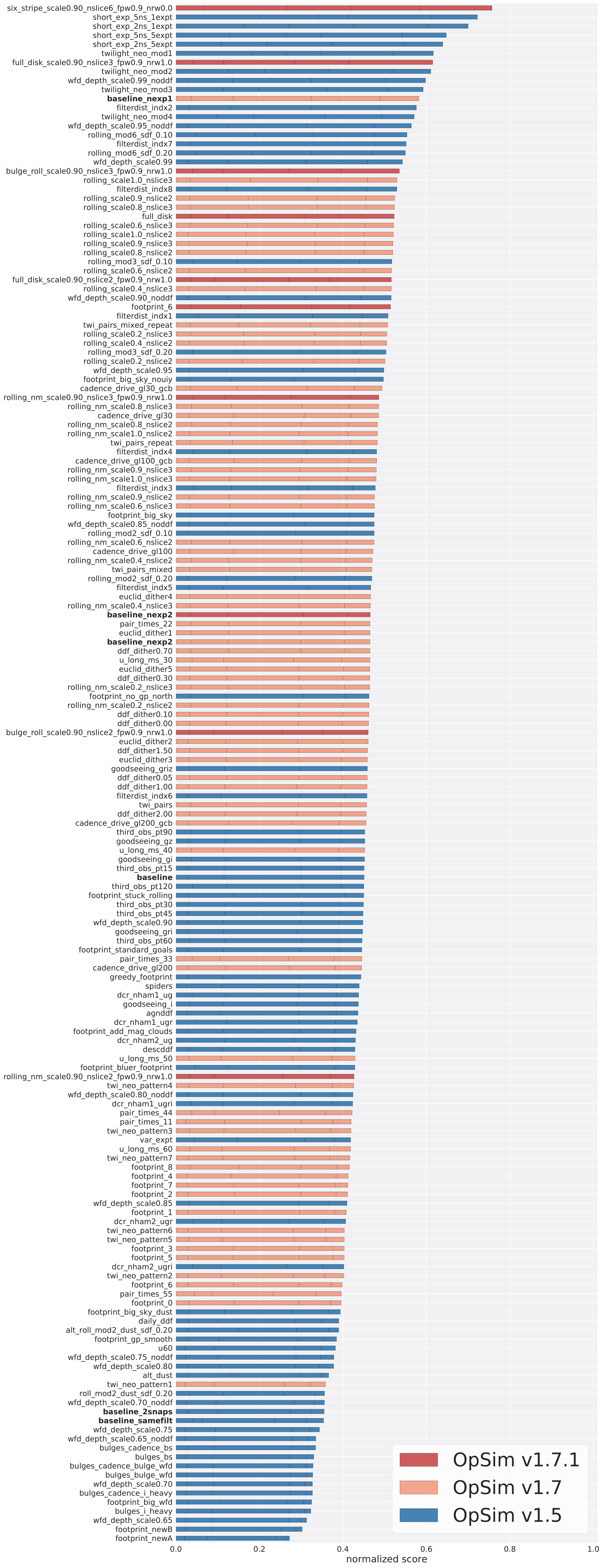}{0.4\textwidth}{ ($b$)}
}
\caption{Bar plot showing the ranking of \opsim s based on our five-fold \fom~for WFD visits (selected as \texttt{proposalId=1}). All simulations from \ovfive,  \ovseven, and \ovsevenone~are included. ($a$) shows the result of our \fom~while ($b$) shows the result after scaling the number of visits in \ovfive~by 7\% to isolate the impact of small differences in survey efficiency associated with the single visit collection strategy (1$\times$30 seconds \emph{vs} 2$\times$15 seconds).
The contribution of each {component} of our \fom~is shown in the same order as in \autoref{fig:barh_wfd}, \autoref{fig:barh_ms_wfd}, and \autoref{fig:barh_ms_all}: \emph{flux change, color, depth, footprint}, and \emph{star count} from left to right. This plot is discussed in
\autoref{ss:v1.7}}\label{fig:barh_v15_v17}
\end{figure*}
\begin{figure*}
\centering
\gridline{
  \fig{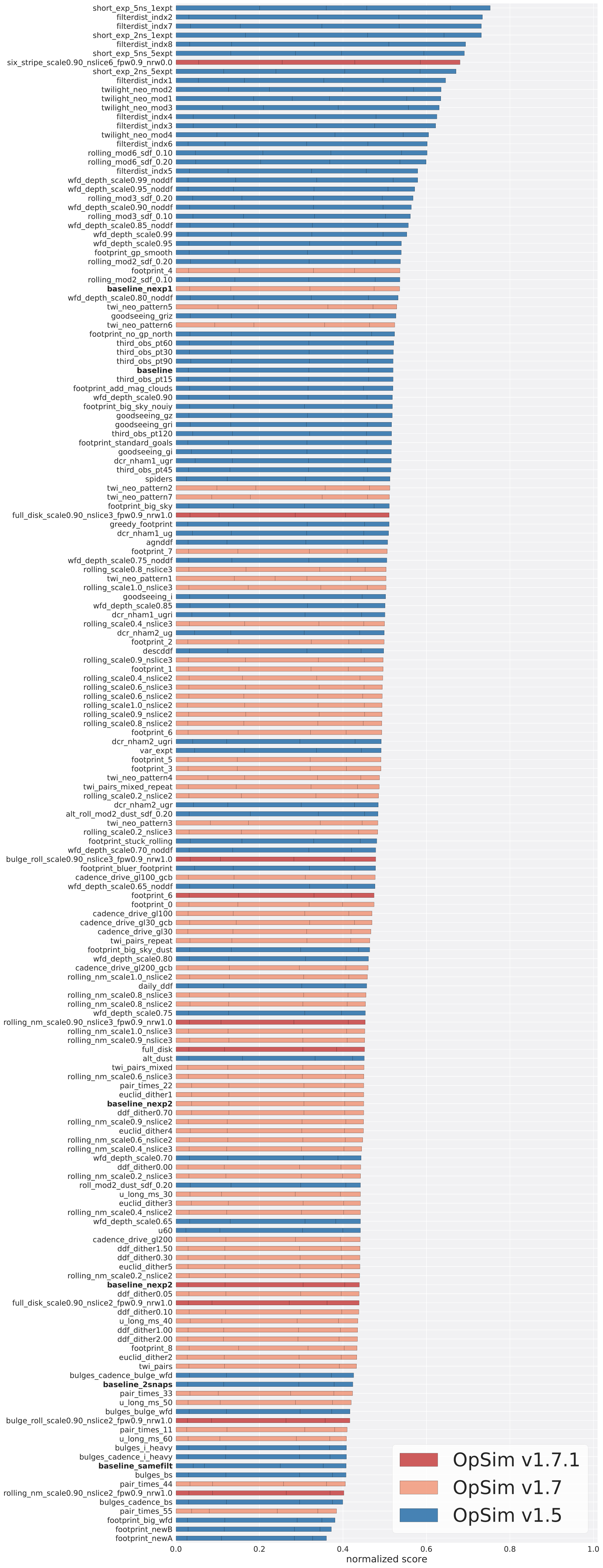}{0.4\textwidth}{($a$)}
  \fig{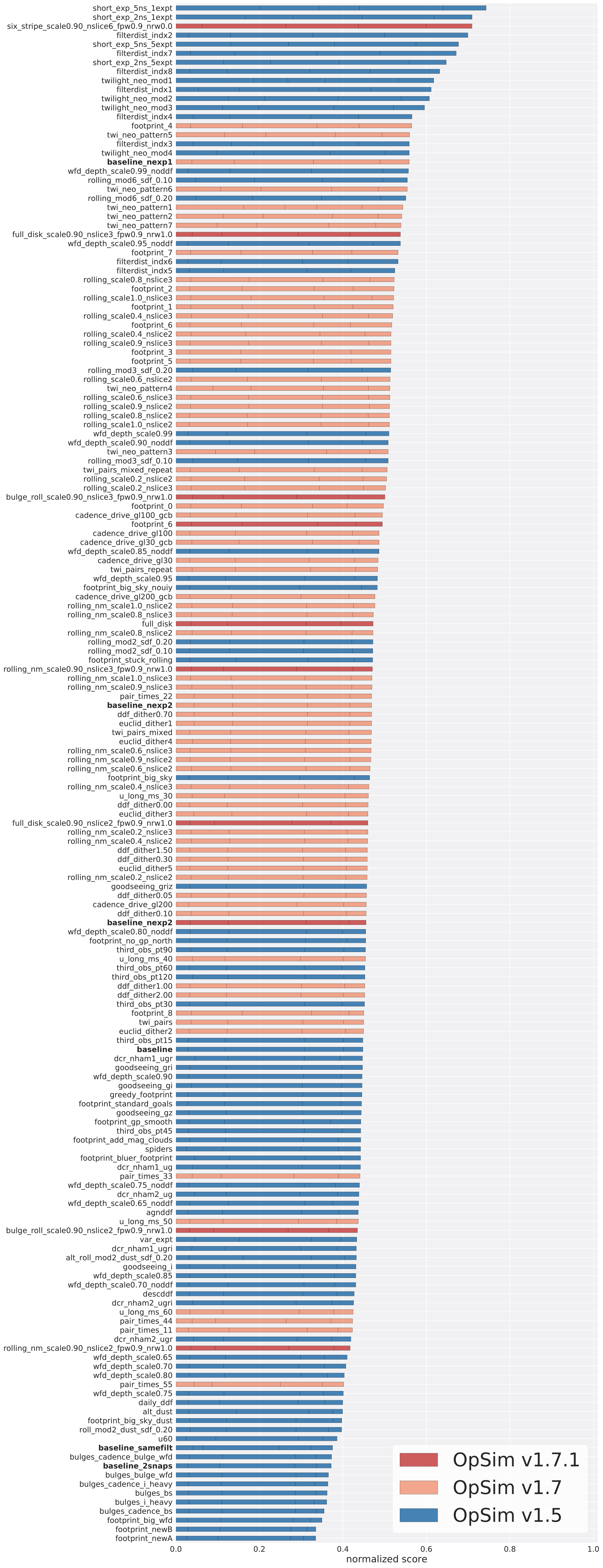}{0.4\textwidth}{($b$)}
}
\caption{As \autoref{fig:barh_v15_v17} but for all regions of the sky that do not correspond to DDFs.   This plot is discussed in
\autoref{ss:v1.7}}\label{fig:barh_v15_v17_all}
\end{figure*}

{ \autoref{fig:barh_v15_v17} and \autoref{fig:barh_v15_v17_all} show our \fom~ for all \opsim s with the three \opsim~versions side by side,  color-coded by \opsim. \autoref{fig:barh_v15_v17} shows the results for observations identified with the WFD survey (\texttt{proposalId=1}), while  
\autoref{fig:barh_v15_v17_all} shows the results for all observations not identified with Deep-Drilling Fields (and thus addressing the impact of the inclusion of the minisurveys in the overall science figures of merit).}

When run on our final \fom, 
\ovfive~leads in general to larger \fom~values (and thus suggests greater scientific yield). 
In \autoref{fig:barh_v15_v17}-$a$, we can observe how almost all \ovfive s~(blue) outperform \ovseven s~(orange), while \ovsevenone~ simulations populate all regions of the chart, with \texttt{six\_stripe\_scale0.90\_nslice6\_fpw0.9\_nmw0.0} outperforming all others. This is a rolling cadence, with six declination stripes as the rolling scheme. This \opsim~performs well on all components of our metrics except the piece that measure flux change (\fom$_\mathrm{tGaps}$ in the same filter) where this \opsim~is outperformed, as discussed in \autoref{sec:timegaps} and \autoref{sec:discussion}, by \opsim s that include short exposures. However, the performance on measuring color (\ie~the number of observations within 1.5 hours in different filters) and the footprint components of the \fom~are sufficient to compensate for this and place the \opsim~at the top.

We suspect that much of the difference between \opsim~versions may be due to the sensitivity of the figures of merit to the total number of observations collected, through the $\sim 7\%$~reduction in the number of observations per field noted above (particularly considering that our footprint \fom s are based on a threshold). 

To test this hypothesis, we {scale down by $7\%$~the number of visits in the calculations of} the \fom$_\mathrm{tGaps}$ and \fom$_\mathrm{footprint}$ elements applied to \ovfive. 
The results of this exercise are shown in panel $b$ of  \autoref{fig:barh_v15_v17} and \autoref{fig:barh_v15_v17_all}.

This in fact does mitigate the almost binary split in performance between the two \opsim~versions seen in panel $a$, although the \texttt{twilight} and \texttt{short}  simulations from version \ovfive~ continue to be at the top. 
Correcting for the $7\%$~depth effect, the \ovsevenone~release also improves relative to \ovfive~ (as expected) and now all nine of the 10 simulations in \ovsevenone~ are in the top 50\%.
{ Indeed, once we control for the overall number of observations, the \ovseven~and \ovsevenone~evaluations tend to populate the upper half of the distribution. However, most of the very highest-performing \opsim s by our \fom s still belong to \ovfive: we note that seven of the top eight-performing \opsim s in \ovfive~include exploration of short exposures (\autoref{fig:barh_v15_v17}-$b$).}

{Inclusion of the minisurveys seems to mitigate slightly the preference for 1$\times$30s exposures, with a handful of \opsim s from \ovseven~now appearing in the top quartile (\autoref{fig:barh_v15_v17_all}-$a$). As with the WFD-only observations, the overall number of exposures seems to explain most of the discrepancy between \ovfive~and the newer releases (\autoref{fig:barh_v15_v17_all}-$b$).}

While the performance across \ovsevenone~ simulations is diverse, \texttt{six\_stripe}\texttt{\_scale0.90}\texttt{\_nslice6}\texttt{\_fpw0.9}-\texttt{\_nmw0.0} stands out as a high throughput observing scheme for our science. 

\section{Conclusion}\label{sec:conclusion}

{Rubin LSST is designed to transform entire fields of astronomy by collecting an unprecedentedly large and rich photometric data set. Yet one of the most exciting promises of LSST is its potential to discover completely novel phenomena, never before observed or predicted from theory. We created a five-fold \fom~that relies on a set of \maf s that assesses the ability of Rubin Observatory LSST to discover novel astrophysical objects, but instead of selecting known anomalies (\eg,  \citealt{tabbysstar}) or theoretically predicted unusual phenomena to benchmark our results, as more commonly done in the field ~\citep{Soraisam20, Pruzhinskaya19, Ishida19, Aleo20,  vafaei2019flexible, martinez20, lochner2020astronomaly, doorenbos2020comparison}  we attempted to remain true to the premise that a true novelty is something that fundamentally cannot be predicted. This exercise is conceptually difficult as by definition we do not know what we are looking for. We can however rely on the completeness of the feature space derived from the survey's data: if all measurable features are exhaustively sampled, anomalies can be detected. 

We thus created a series of \maf s and \fom s that measure the completeness in the space of observables derived from LSST data. Completeness to color and magnitude (and their evolution) was probed by measuring the number of observations and time gaps between observations in pairs of different filters and in the same filter (respectively). We scaled a survey quality by the survey's sky coverage, choosing to benchmark this component of the metric to a fiducial implementation of LSST, \texttt{baseline\_v1.5}, and by the number of objects observed, scaling the footprint itself by the number of stars in each field. These metrics were then summed into a single \fom. Finally, since the \fom~so far assembled largely relies on number of observations, an \fom~ element was needed that considers the \emph{quality} of the observations. For this we added \fom$_\mathrm{depth}$~to measure the LSST 10-year stacks magnitude depth, penalizing for example \opsim s that include short-exposure observations if these take time from high-quality, deep observations, but only in this case. Proper motion considerations are reserved for paper II.}
 
{While the main purpose of this paper is to conceptualize a non-parametric way to explore a survey's potential for anomaly detection, these considerations will ultimately need to be applied to current and future Rubin LSST candidate strategies. To illustrate how this can be done, we performed the comparisons for recent suites of simulations.}

We identified some high-performing families within \ovfive~ and justified their high rank as measured by our \fom~(\autoref{ss:mainsurvey} and \autoref{ss:minisurveys}). {Generally, families of \opsim~ that maximize the diversity of the observations (in terms of time gaps, footprint and exposure time) seem to be preferred, but there is considerable variation within each family.} 

{To first order, as expected, the mini-surveys seem to be led by footprint considerations. Since fundamentally the allocation of observations to minisurvey regions is a zero-sum game, we point out here that there are high-performing \opsim s for the minisurveys that do not dramatically impact the science obtained in the main-survey - so allocating a modest number of exposures to the minisurveys does not seriously impact the scientific goals of the main survey.}

We briefly inspected the most recent (at the time of writing) versions of the \opsim: \ovseven~and \ovsevenone~and found that their performance is impacted, in general, by collecting exposures in two snapshots (2$\times$15 seconds \emph{vs} 1$\times$30 seconds). 

However, even correcting for this, some families of \ovfive~simulations are the best performers for our science case: namely those that provide visibility into additional time scales by adding short exposures to the observing plan, but planning them when long exposures are unfeasible, so that they do not come at the cost of an overall loss of survey depth (\eg, \texttt{twilight}). {We point out that any extension of the feature space is advantageous to the discovery of true-novelties, and thus 
we are not bound to the minimum allocation of short exposures required for other goals (such as cross-calibration of LSST to external catalogs with brighter saturation limits; \eg~\citealt{gizis19}); those considerations are beyond the scope of this paper.}

{A further comment on the issue of 1$\times$30 vs 2$\times$15 seconds is in order.}
While we see the effects of increased survey efficiency in our metrics, it should be emphasized that none of our metrics include considerations on the impact of this choice on image quality or on the 
{capability to open up {\it intra-visit} timescales by treating the two exposures in a visit separately. {Combining the impact of Rubin's LSST data volume with visibility into short-time scales is potentially transformational for rare phenomena
(like \eg, relativistic explosions, see \autoref{fig:phasespace}).}}
We note, however, that any analysis based on the individual snaps that make the 30 seconds exposure would require custom pipelines.

Yet a comprehensive discussion of the detailed reasons why a specific \opsim~achieves a certain performance is beyond the scope of this paper. We encourage instead the use of our metrics to evaluate existing and new \opsim~ to implement an LSST survey that maximizes the throughput of Rubin Observatory in its four science pillars with particular care to the discovery novelties, that has the potential to advance or transform all of these fields. 

The code on which this analysis is based is available in its entirety in a dedicated GitHub repository\footnote{\url{https://github.com/fedhere/LSSTunknowns}}. 

\begin{acknowledgments}

This paper was created in the nursery of the Vera C. Rubin Legacy Survey of Space Time Science Collaborations\footnote{\url{https://www.lsstcorporation.org/science-collaborations}} and particularly of the Transient and Variable Star Science Collaboration\footnote{\url{https://lsst-tvssc.github.io/}} (TVS SC) and Stars, Milky Way, and Local Volume Science Collaboration\footnote{\url{https://milkyway.science.lsst.org/}} (SMWLV SC).

The authors acknowledge the support of the Vera C. Rubin Legacy Survey of Space and Time TVS SC and SMWLV SC that provided opportunities for collaboration and exchange of ideas and knowledge.
The authors are thankful for the support provided by the Vera C. Rubin Observatory \maf~team in the creation and implementation of \maf s.
The authors acknowledge the support of the LSST Corporations that enabled the organization of many workshops and hackathons throughout the cadence optimization process through private fundraising.

The authors thank Dr. Edward Ajhar, who emphasized the importance of an evaluation of the effectiveness of the Rubin survey strategy in the discovery of unknown phenomena at the 2019 LSST (Rubin) Project Community Workshop.

We used the following packages:
\begin{itemize}
\item{\texttt{python} including}
    \begin{itemize}
   
   \item \texttt{numpy} \citep{harris2020array}
      \item \texttt{maplotlib} \citep{matplotlib}
      \item \texttt{scikit-learn} \citep{JMLR:v12:pedregosa11a}
     \item{\texttt{pandas}} \citep{pandas}
      \item \texttt{seaborn} \citep{seaborn}
    \end{itemize}
\end{itemize} 
\begin{itemize} 
\item{the \texttt{glasbey} package to generate maximally separable colors. \citep{glasbey2007colour}}
\item{Webplot Digitizer \citep{digitizer}}
\item{\texttt{rsmf} (right-size my figures)\footnote{\url{https://github.com/johannesjmeyer/rsmf}}}
\item{\texttt{d3.js} \footnote{\url{https://d3js.org}} to create spatial selection tools and interactive radar/parallel plots.}
\end{itemize}

\end{acknowledgments} 

\newpage
\appendix
\section{Interactive tools}\label{appendix}
\counterwithin{figure}{section}
\counterwithin{figure}{section}
We provide three interactive tools that support the analysis performed in this work.

\begin{figure*}
\gridline{
  \fig{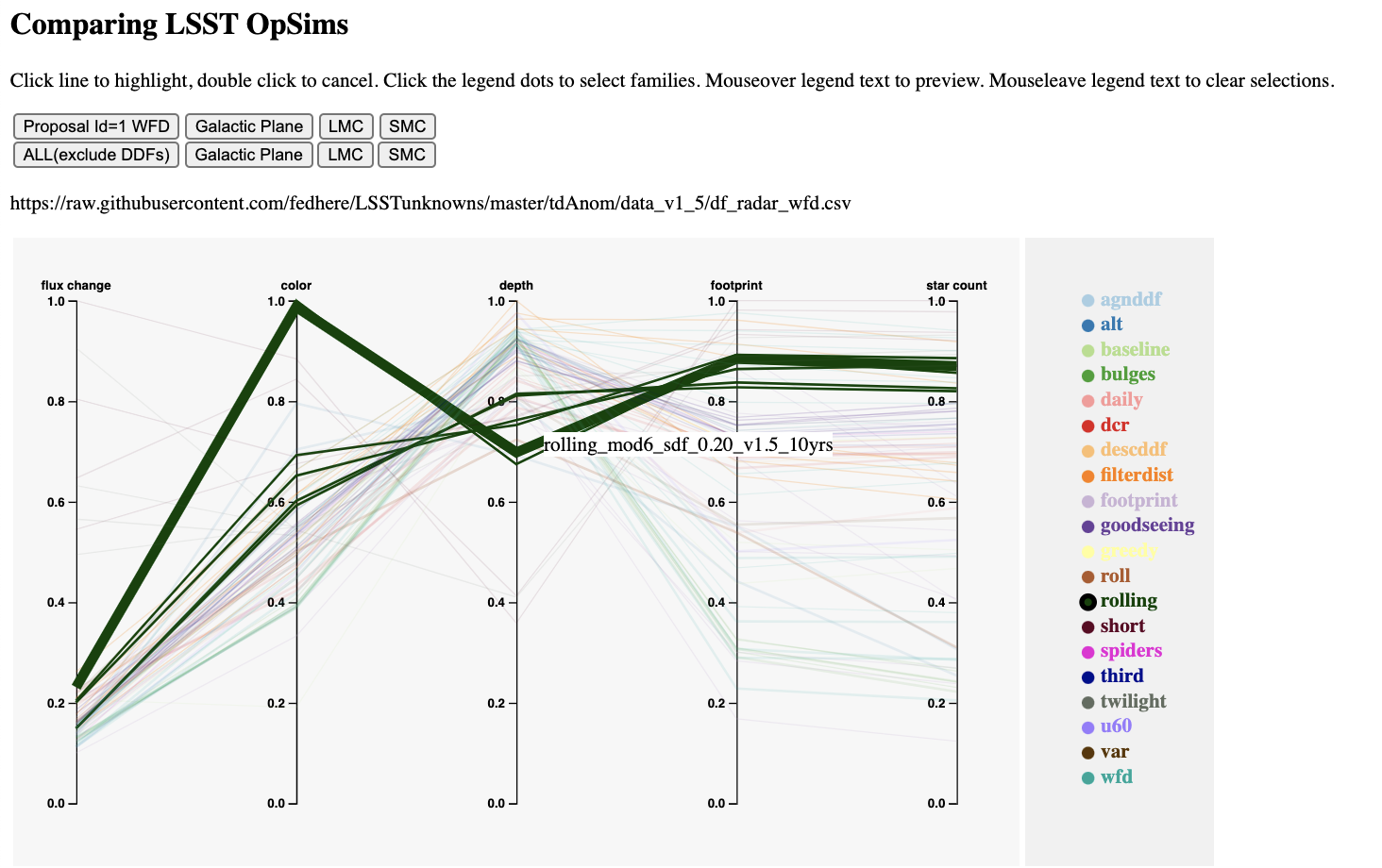}{0.5\textwidth}{($a$)}
  \fig{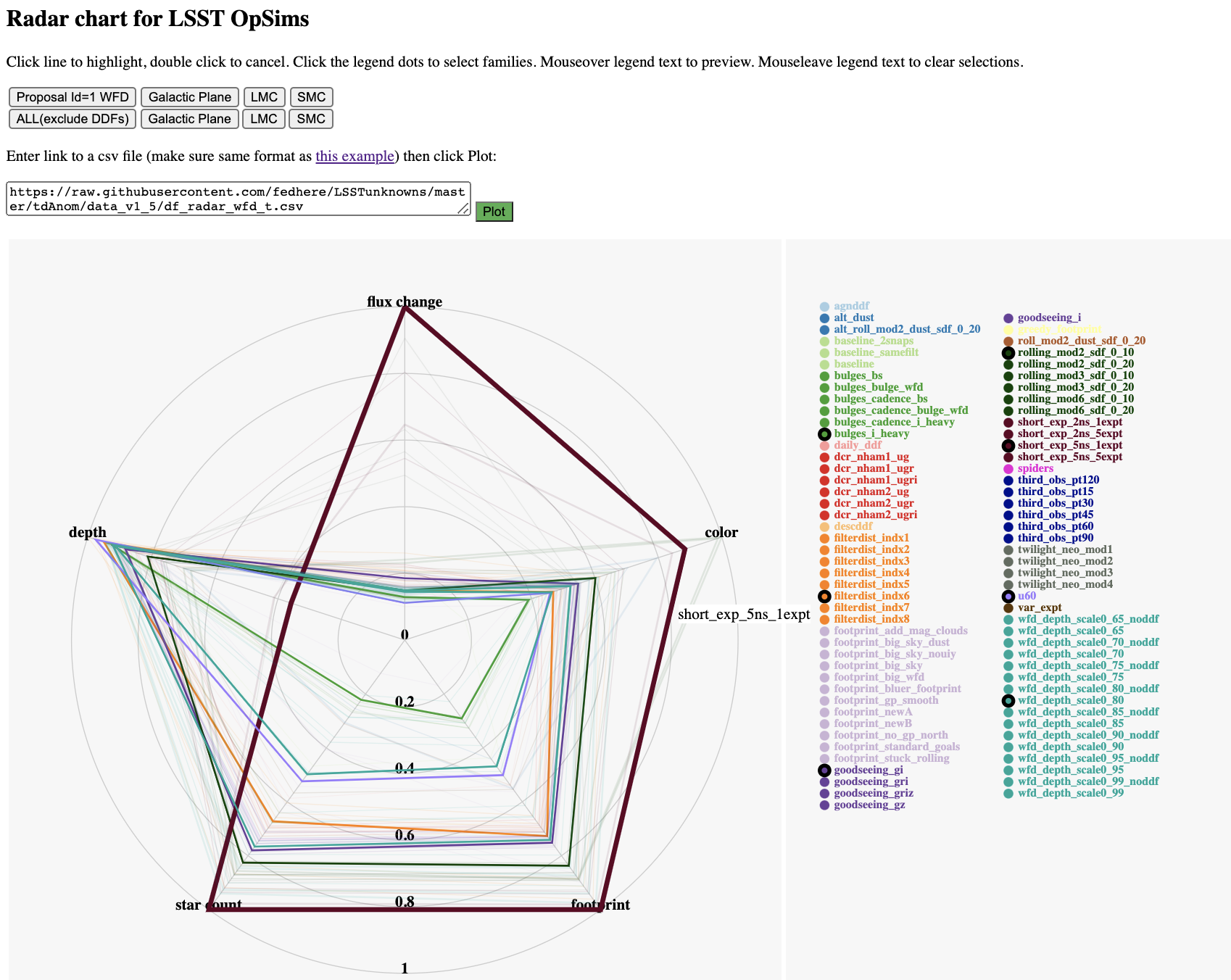}{0.38\textwidth}{($b$)} 
  }
\caption{Snapshots of the interactive versions of two kinds of summary plots: a parallel coordinate plot ($a)$ and a radar plot ($b$). These interactive visualizations are made available to the reader to explore our metrics or load their own. See \autoref{appendix}.} \label{fig:widgets}
\end{figure*}

We make  javascript-D3 \citep{d3} interactive versions of two synoptic visualization of the results of our \fom~available: a parallel coordinate plot and a radar plot. 

The parallel coordinate plot\footnote{\url{https://xiaolng.github.io/widgets/parallel.html}} (\autoref{fig:widgets} panel $a$) allows the user to follow the performance of an \opsim~across components of the \fom. Toggling between families of \opsim s to highlight the \opsim s within, while keeping all other \opsim s in the background, the user can easily identify ``standout'' \opsim s by \fom~element. By selecting the ``cumulative'' option the viewer can follow the evolution of an \opsim~across components of the \fom~while retaining information about the overall performance.

The radar plot\footnote{\url{https://xiaolng.github.io/widgets/radar.html}} (Figure \ref{fig:widgets} panel $b$, also discussed in \autoref{sec:discussion}) is a synoptic visualizations that maps multiple elements of a \fom~to a polygon, with the distance from the center of the polygon representing the result of the \fom~element. It allows the user to visualize tension between component of the \fom~as well as the overall quality of an \opsim, which maps to the area of the polygon. It is however hard to include many \opsim s in the same radar plot without compromising readability. This widget allows the reader to toggle between \opsim s, which are color-coded by family. 

In both widgets is also possible to select the survey or sky area that the user wants to inspect (\eg, WFD, or LMC, etc, see \autoref{ss:featurespace}). 

While both widgets come pre-loaded with the metrics developed in this paper, the user can easily visualize their own metrics by uploading a comma-separated-value format file with the result of the \maf s containing the following columns:
\texttt{db} (the database name), \texttt{m$_1$} (numerical value for the first element of the metric), \texttt{m$_2$}  (numerical value for the second element of the metric), ... , \texttt{m$_n$}  (numerical value for first the last element of the metric).

We offer a \texttt{python} based widget to select regions of sky based on a specific pixelization (\eg, healpix) which was used to select the Galactic Plane, LMC, and SMC regions in \autoref{ss:minisurveys}. This tool is available in a dedicated GitHub repository\footnote{\url{https://github.com/xiaolng/healpixSelector}} as a \texttt{jupyter} notebook and interactive webtool \citep{li2021}.

\begin{figure*}
\centering
\includegraphics[width=0.5\textwidth]{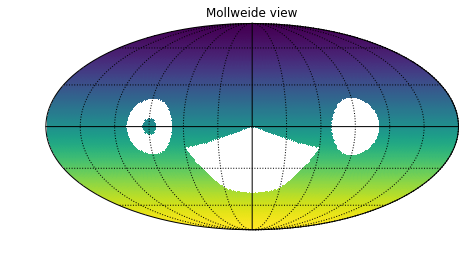}
\caption{Healpixel-based selection of sky areas performed with our widget: convex, concave, and hollow regions can be selected on multiple spatial projections (Mollweide shown). See  \autoref{appendix}.} \label{healpixselect}
\end{figure*}



\end{document}